\documentclass[twocolumn,showpacs,preprintnumbers,amsmath,amssymb,floatfix]{revtex4}

\usepackage[latin1, ansinew]{inputenc}
\usepackage{graphicx}
\usepackage{dcolumn}
\usepackage{bm}

\begin{document}

\title{Force chains and contact network topology in packings of elongated particles}

\author{Emilien Az\'ema and Farhang Radja\"\i}

\affiliation{LMGC, Universit\'e Montpellier 2, CNRS, Place
Eug\`ene Bataillon, 34095 Montpellier cedex 05, France.}

\email{emilien.azema@univ-montp2.fr \\ franck.radjai@univ-montp2.fr}

\date{\today}

\begin{abstract}
By means of contact dynamic simulations, we investigate the 
contact network topology and force chains in two-dimensional packings  
of elongated particles modeled by rounded-cap rectangles.  
The morphology of large packings of elongated particles in 
quasistatic equilibrium is complex due to the combined effects of 
local nematic ordering of the particles and orientations of contacts 
between particles. We show that particle elongation affects  
force distributions and force/fabric anisotropy via various 
local structures allowed by steric exclusions and the requirement of 
force balance. As a result, the force distributions become 
increasingly broader as particles become more elongated. Interestingly, 
the weak force network transforms from a passive stabilizing agent 
with respect to strong force chains   
to an active force-transmitting network for the whole system. 
The strongest force chains are carried 
by side/side contacts oriented along the principal stress direction.
\end{abstract}

\maketitle

\section{Introduction}

Most remarkable properties of granular materials are closely 
related to their specific disorder induced essentially by steric exclusions and 
the force balance condition for each particle. 
The broad and strongly inhomogeneous distribution of contact forces, 
as a hallmark of granular disorder, has been a subject of extensive investigation  \cite{Liu1995a,Radjai1996,Jaeger1996,Mueth1998a,Lovol1999,
Bardenhagen2000,Silbert2002,Silbert2006,Eerd2007,Richefeu2007}. 
In spite of particle mobility and disorder, granular materials exhibit a 
finite shear strength due to a genuine anisotropic two-phase organization of the contact network  
involving strong force chains propped by weak forces \cite{Radjai1998}. 

The robustness of these micro-structural features with respect to particle 
geometry and interactions has been addressed only recently by discrete-element 
numerical simulations. For example, it is found that in highly polydisperse systems 
the force chains are mainly captured by large particles so that the shear strength 
of a noncohesive granular material is practically independent of particle size 
distribution \cite{Voivret2009}.  
As another important example, a parametric study shows that 
when the particles interact by both sliding friction and high rolling 
resistance at their contacts, the nature of the weak network 
is affected by the formation of  columnar structures which do not need to be
propped by a particular class of weak contacts \cite{Estrada2008}. 

The particle shape is another major characteristic of granular material.   
Most applications of real granular materials involve some degree of deviation 
with respect to simple circular or spherical shapes often 
used in simulations by the discrete-element method. 
While the numerical treatment of large packings of complex particle 
shapes  was until very recently  out of reach due to  
demanding computational resources, there is presently considerable 
scope for the numerical investigation of 
complex granular packings. This is not only due to the enhanced computer 
power and memory but also because during 
more than two decades of research in this field, many properties of  
granular media have been investigated for model packings composed of 
circular and spherical particle shapes. Such properties provide thus a 
rich guideline for the analysis of specific behaviors arising from particle 
geometry.      

Systematic studies of particle shape dependence in granular materials 
have been recently reported for polygonal/polyhedral \cite{mirghasemi2002,Azema2007,Azema2009,Estrada2011,Azema2011},  
elliptical/ellipsoidal \cite{C.R.A.Abreu2003,Donev2007,Wouterse2007,Azema2010} and non-convex shapes \cite{Saint-Cyr2011}.       
The force chains are found to be reinforced 
in packings of polygonal and polyhedral particles leading to 
enhanced shear strength \cite{Azema2007,Zuriguel2007,Azema2010}.
The effect of shape elongation was investigated for packings of rectangular-shaped  
particles deposited under gravity \cite{Hidalgo2009}. The preparation under gravity 
has strong influence on the particle orientations and thus on the force 
distributions. On the other hand, a systematic study of the shear 
behavior of 2D packings of rounded-cap rectangles (RCR) under    
homogeneous boundary conditions indicates that 
the shear strength increases with elongation whereas the 
packing fraction varies unmonotonically \cite{Azema2010}, as also 
found for packings of ellipsoidal shapes \cite{Donev2007,Wouterse2007}. 
In all reported cases, the networks resulting from various 
shapes appear to be highly complex and hardly amenable 
to simple statistical modeling.        

In this paper, we use contact dynamics simulations to investigate the 
contact and force networks in sheared granular packings of RCR particles 
with increasing aspect ratio in 2D. We focus more specifically on the organization of 
the contact force network in correlation with the fabric anisotropy described in terms of 
branch vectors joining particle centers. Our data reveal a bimodal force network  
as in disk packings but with qualitatively different roles of fabric and force 
anisotropies. This behavior involves a short-range nematic ordering of 
the particles with side/side contacts that capture stronger 
force chains. On the other hand, the friction mobilization is shown 
to be anisotropic and it plays a major role for the stability of 
elongated particles.   

In the following, we first briefly describe the numerical procedures, which 
are essentially the same as  those reported in \cite{Azema2010}. Then, we analyze 
the branch vectors and their correlations with the contact forces. Finally, we present 
a detailed analysis of the partial stresses and fabric anisotropies 
sustained by force sub-networks. We conclude with salient results of this 
work its possible prospectives.        

\section{Model description and numerical simulations}
\label{procedure}

The simulations were carried out by means of the contact dynamics (CD) method with 
irregular polyhedral particles. 
The CD method is a discrete element approach for the simulation of nonsmooth granular dynamics with contact laws expressing  
mutual exclusion and dry friction between particles without elastic or viscous regularization 
\cite{Moreau1994,Radjai1997,Jean1999,Moreau2004,Dubois2006,Richefeu2007,Radjai2009,Radjai2011}. Hence, this method is particularly adapted for the simulation of perfectly rigid particles.
Nonsmoothness refers to various degrees of discontinuity in velocities arising in a system of rigid particles.     
In this method, the equations of motion for each particle are formulated as differential inclusions in which
velocity jumps replace accelerations \cite{Moreau1994}. 
The unilateral contact interactions and Coulomb friction law are treated as  complementarity relations 
or set-valued contact laws.  The time-stepping scheme is implicit but requires explicit determination of the  
contact network. Due to implicit time integration, inherent in the CD method, this scheme is unconditionally stable. 

At a given step of evolution, all kinematic constraints implied by lasting  contacts and the possible 
rolling of some particles over others are simultaneously taken into account, together with  the equations 
of dynamics, in order to determine all velocities and contact forces in the system. 
This problem is solved by an iterative process  pertaining to the non-linear Gauss-Seidel method which consists
of solving a single contact problem, with other contact forces being treated as known, and iteratively 
updating the forces and velocities until a convergence criterion is fulfilled. 
The iterations in a time step are stopped when the calculated contact forces are stable 
with respect to the update procedure.
To check convergence we thus use the relative variation of the mean contact force between two successive iterations.
We require this relative variation to be below a given value which sets the precision of the calculation.   
In this process, no distinction is made between smooth evolution of a system of 
rigid particles during one time step and nonsmooth 
evolutions in time due to collisions or dry friction effects.
The uniqueness of the solution at each time step is not guaranteed by CD method 
for perfectly rigid particles. However, by initializing each step of calculation with the forces calculated  
in the preceding step, the set of accessible solutions shrinks to fluctuations which are basically below the numerical resolution. 
In this way, the solution remains close 
to the present state of forces.

For our simulations, we used the LMGC90 which is a multipurpose software developed in Montpellier, capable of modeling a collection of deformable or undeformable particles of various shapes (spherical, polyhedral, or polygonal) by different algorithms
\cite{Dubois2006,Radjai2011}.

\subsection{Simulation of RCR particles}

\begin{figure}
\includegraphics[width=4cm]{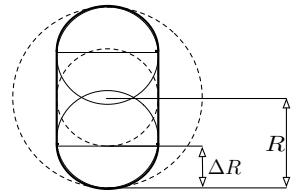}
\caption{Shape of a Rounded-Cap Rectangle (RCR). 
 \label{sec:numerical_procedure:def_jonc}}
\end{figure}

We model the RCR particle as a juxtaposition of two half-disks of radius $R'$ 
with one rectangle of length $L$ and width $2R'$; see 
Fig. \ref{sec:numerical_procedure:def_jonc}. 
The shape of a RCR particle is a circle of radius $R'$ for $L=0$. The aspect ratio 
$\alpha = (L+2R')/(2R')$  is $1$ in this limit and increases with $L$ for a fixed value of $R'$.  
In this paper, we use an alternative parameter describing the deviation of the particle shape 
from a circle. Let $R$ be the radius of the circle circumscribing the particle. 
We have $R=L/2+R'$. The radius $R'$ is also that of the inscribed circle. 
Hence, the deviation from a circular shape can be characterized by $\Delta R = R-R' = L/2$. 
We use the dimensionless parameter $\eta$ defined by 
\begin{equation}
\eta = \frac{\Delta R}{R} = \frac{\alpha-1}{\alpha}.
\label{Def_eta}
\end{equation}
It varies from $\eta=0$, for a circle,  
to 1 corresponding to a line. We will refer to $\eta$ as the {\em elongation} 
parameter as in rock mechanics \cite{Folk1974}. 

\begin{figure}
\includegraphics[width=2.5cm]{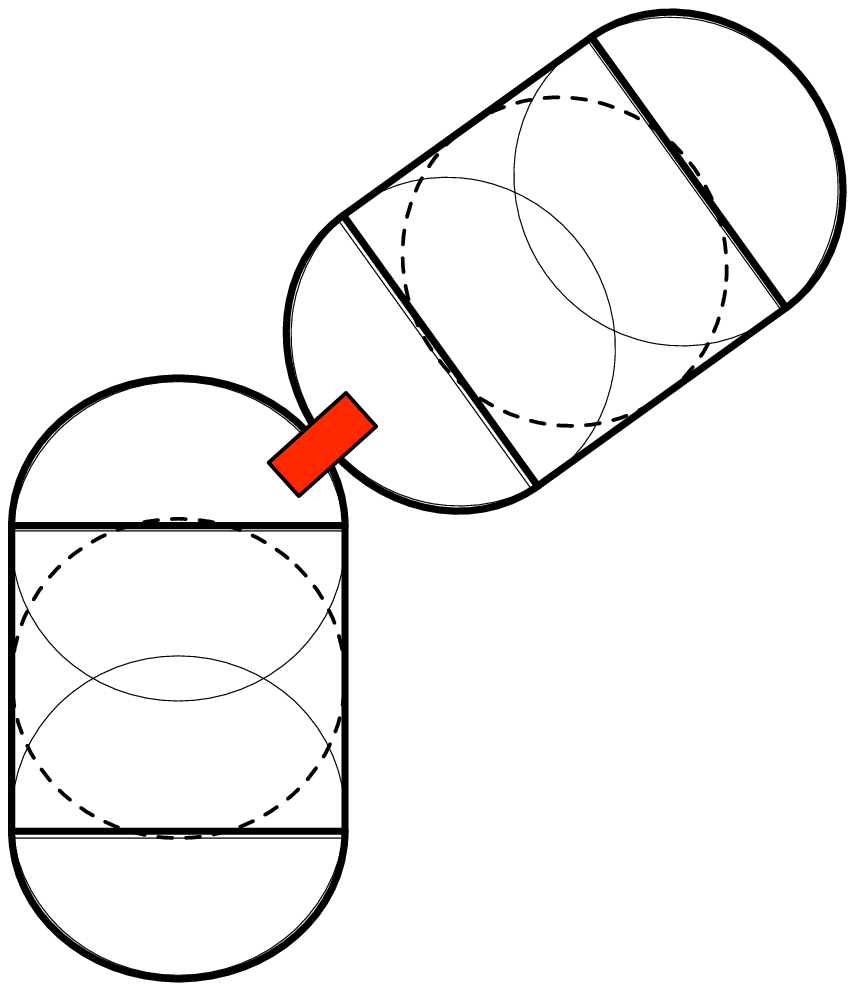}
\includegraphics[width=2.5cm]{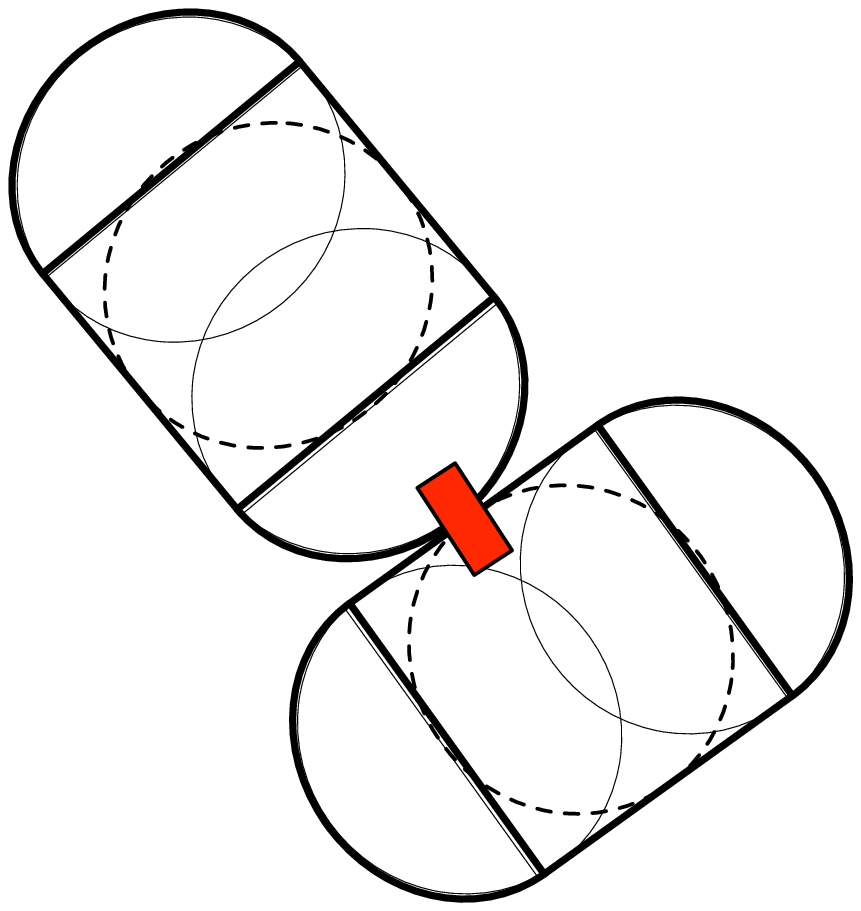}
\includegraphics[width=2.5cm]{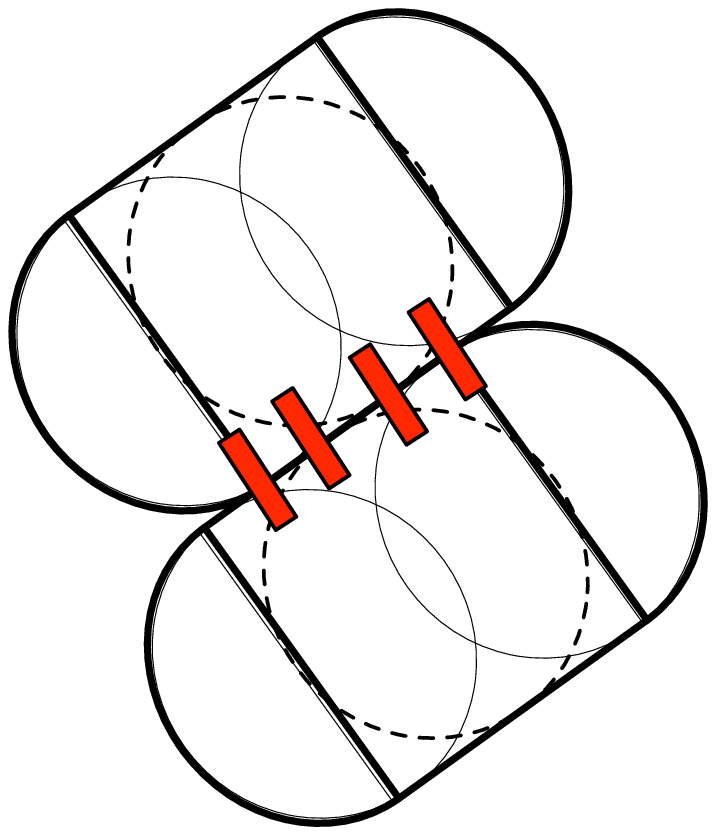}
\caption{Representation of cap-to-cap, cap-to-side  
 and side-to-side contact and they will be referred  as  $cc$ contacts, $cs$ contacts and $ss$ contact, respectiveley.\label{sec:numerical_procedure:type_contacts}}
\end{figure}

The contacts between RCR particles belong to different categories, 
namely cap-to-cap ($cc$), cap-to-side ($cs$) and side-to-side ($ss$); 
see Fig. \ref{sec:numerical_procedure:type_contacts}. 
Side-to-side contacts results from contacts between two rectangles as well as
two contacts resulting from cap-to-side. In the CD method the case of side-to-side contacts 
for rectangular particle is represented by two points. Hence, for RCR particles, 
$ss$ contact is composed of four point contacts : two points due to the 
rectangle-rectange interface and two points due to the $cs$ contacts. 
In the iterative procedure of determination of the contact forces and velocities, 
the  points representing the contact between two particles
are treated as independent points but the resultant of the calculated forces 
are attributed to the contact with its application point located on the contact plane.   

The detection of line contacts between rectangles was implemented through 
the so-called {\it shadow overlap method} devised initially by 
Moreau \cite{Saussine2006,Radjai2011} for polygons. The reliability and robustness of this method 
have been tested in several 
years of previous applications to  granular materials 
\cite{Nouguier-Lehon2003,Azema2006,Saussine2006,Azema2007,Azema2008,Azema2009,Radjai2011}. 
This detection procedure is fairly rapid and allows us to simulate large samples 
composed of RCR particles. 

\subsection{Packing preparation and bi-axial test}
We prepared 8 different packings of $13000$ RCR particles 
with $\eta$ varying  from 0 to $0.7$ by steps of $0.1$. 
The radius $R$ of the 
circumscribing circle defines the size of a RCR particle. 
In order to avoid long-range ordering in the limit of small values 
of $\eta$, we introduce a size polydispersity by taking $R$ in 
the range $[R_{min} , R_{max} ]$ with $R_{max} = 2R_{min}$ 
with a uniform distribution in particle volume fractions.

All samples are prepared according to the same protocol. A dense 
packing composed of disks ($\eta=0$) is first constructed by means of a 
layer-by-layer deposition model based on simple geometrical rules 
\cite{Bratberg2002,Taboada2005,Voivret2007}. 
The particles are deposited sequentially on a substrate. Each new particle
is placed at the lowest possible position at the free surface as a 
function of its diameter. This procedure leads to a random close packing
in which each particle is supported by two underlying particles and
supports one or two other particles. For $\eta > 0$, the same packing is used 
with each disk serving as the circumscribing circle of a RCR particle.   
The RCR particle is inscribed with the given value of $\eta$ and random orientation 
in the disk.

\begin{figure}
\includegraphics[width=8.5cm]{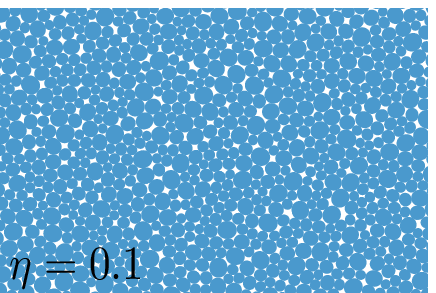}
\includegraphics[width=8.5cm]{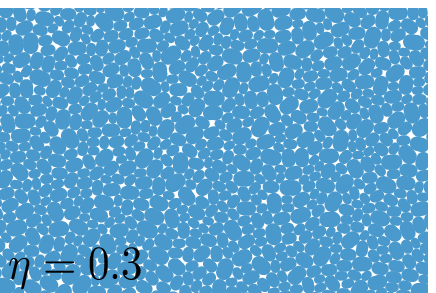}
\includegraphics[width=8.5cm]{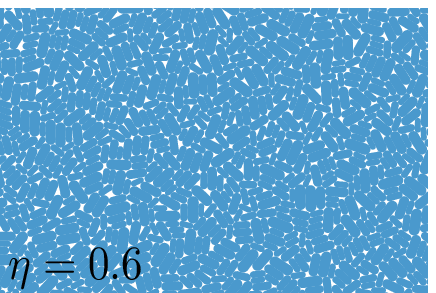}
\caption{Examples of the generated packings at the initial state.  \label{map_ini}}
\end{figure}

Following this geometrical process, the packing is compacted by isotropic 
compression inside a rectangular frame of dimensions $l_0 \times h_0$ in which 
the left and bottom walls are fixed, and the right and top walls are subjected to 
a compressive stress $\sigma_0$. 
The gravity $g$ and friction coefficients $\mu$  between particles 
and with the walls are set to zero during the compression in order to
avoid force gradients and obtain isotropic dense packings. 
Fig. \ref{map_ini}  displays snapshots of the packings for 
several values of $\eta$ at the end of isotropic compaction. 

The isotropic samples were sheared by applying a downward displacement on 
the top wall at constant velocity for a constant confining stress acting on the 
lateral walls; see http://cgp-gateway.org/ref010 for video samples. 
During shear, the friction coefficient $\mu$ between particles was set to 0.5 and to 
zero with the walls. The strain rate was low so that the shearing is basically of quasi-static nature. 
The internal angle of friction $\varphi$ at every stage of shearing is given by 
\begin{equation}
\sin \varphi = \frac{q}{p} = \frac{\sigma_1-\sigma_2}{\sigma_1+\sigma_2},
\label{eq:qp}
\end{equation}
with $\sigma_1>\sigma_2$ are the principals values of the
stress tensor $\bm \sigma$. $\sin \varphi$ increases with 
shear strain and saturates to a constant value corresponding to 
the critical state which is a state independent of the initial configuration 
of the packing. The critical-state value of $\sin \varphi$ represents the 
shear strength of the packing and it increases linearly with $\eta$ 
from $ \sim 0.3$ (for $\eta=0.0$) to $\sim 0.51$ (for 
$\eta=0.7$) \cite{Azema2010}.

In granular media, the expression of stress tensor $\bm \sigma $ in the volume $V$ 
is an arithmetic mean involving the 
branch vectors $\bm \ell^c$ (joining the centers of  the two touching particles) 
and contact force vectors $\bm f^c$ at contact $c$, and it is given by \cite{Moreau1997,Staron2005}:
\begin{equation}
{\bm \sigma } =   \frac{1}{V}  \sum_{c \in V} f_{\alpha}^c \ell_{\beta}^c, 
\label{eq:sigma}
\end{equation}
For the analysis of stress transmission from a particle-scale viewpoint 
we need a statistical description of these quantities. 

\begin{figure}
\includegraphics[width=6cm]{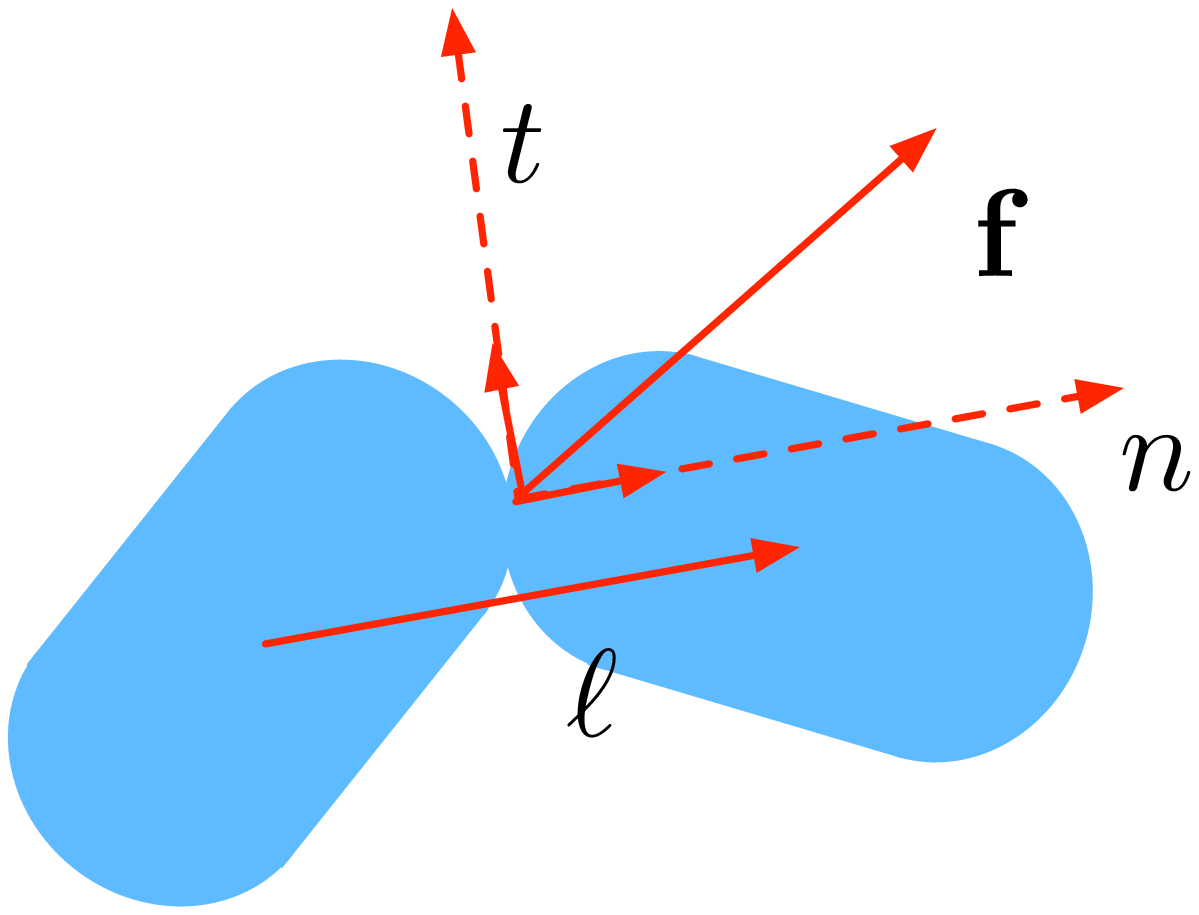}
\caption{Local contact frame $(\bm n,\bm t)$ \label{sec:numerical_procedure:frame}}
\end{figure}

A common approach used by various authors is to express branch vectors and 
contact force orientations
in terms of the contact direction, i.e. in the local {\it contact frame} $(\bm n, \bm t)$, 
where $\bm n$ is the unit vector perpendicular to the contact 
plane, and $\bm t$ is an orthonormal unit vector oriented along the tangential force; 
see figure \ref{sec:numerical_procedure:frame}. The components  
of the branch vector and contact force are expressed in the following frame: 
\begin{equation}
\left\{
\begin{array}{lcl}
\bm \ell &=&  \ell_n \bm n + \ell_t \bm t,\\
\bm f &=& f_n \bm n + f_t \bm t,
\end{array}
\right.
\label{eq:n}
\end{equation}
where $\ell_n$ and $\ell_t$ are the normal and tangential components of the branch vectors, and 
$f_n$ and $f_t$ the normal and tangential components of the contact force.
Remark that only for disks or spherical particles we have $\bm \ell = \ell \bm n$ 
where $\ell$ is the length of the branch vector.

In the following we study the shapes of the normal and tangential force and branch distributions
in the residual state.

\section{Distributions of contact forces and branch vectors}
\label{sec:distribution}

A specific feature of the contact network of a packing of 
elongated particles is that the length $\ell$ of branch vectors 
strongly varies throughout the network depending on the relative 
particle orientations. From the definition of 
$\eta$ (Eq. (\ref{Def_eta})) and for given values of $R_{min}$ and $R_{max}$, it is 
easy to see that 
\begin{equation}         
\frac{\ell}{R_{max}} \in \left[2 \frac{R_{min}}{R_{max}} (1-\eta), 2 \right]
\label{eqn:ellR}
\end{equation}  
In our simulations, since $R_{min}/R_{max}=0.5$, we 
have $(1-\eta)R_{max} \leq \ell \leq 2 R_{max}$. With increasing elongation $\eta$, 
the range of $\ell$ becomes significant and its statistics can be used as a meaningful 
characterization   of the texture as a function of $\eta$. On the other hand, the correlation of 
$\ell$ with the total reaction force $f$ between neighboring particles seems to be a 
good descriptor of the organization of forces for particles of non circular shape. 
The branch vectors are also important as they enter the expression of the 
stress tensor given by Eq. (\ref{eq:sigma}).   
In Ref. \cite{Azema2010}, a different point of view was adopted: the contact 
forces were projected along and perpendicular to the branch vectors and their statistics 
were investigated. The same framework was used for the decomposition of the total 
stress tensor. Here, we focus on the distribution of contact forces 
and their correlation with the branch vector 
as $\eta$ is increased.  

\subsection{Contact forces and friction mobilization}
\begin{figure}
\includegraphics[width=8cm]{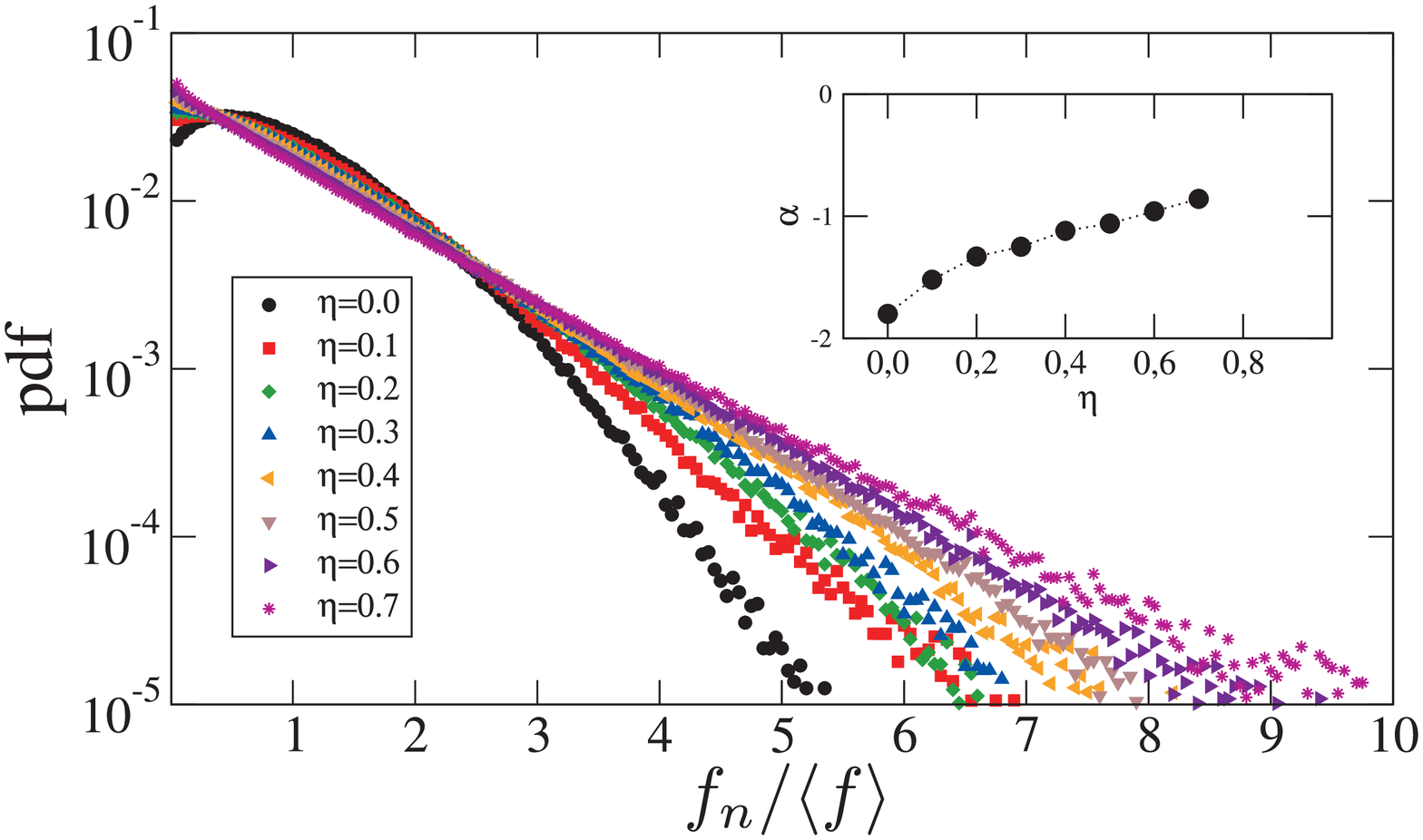} (a)
\includegraphics[width=8cm]{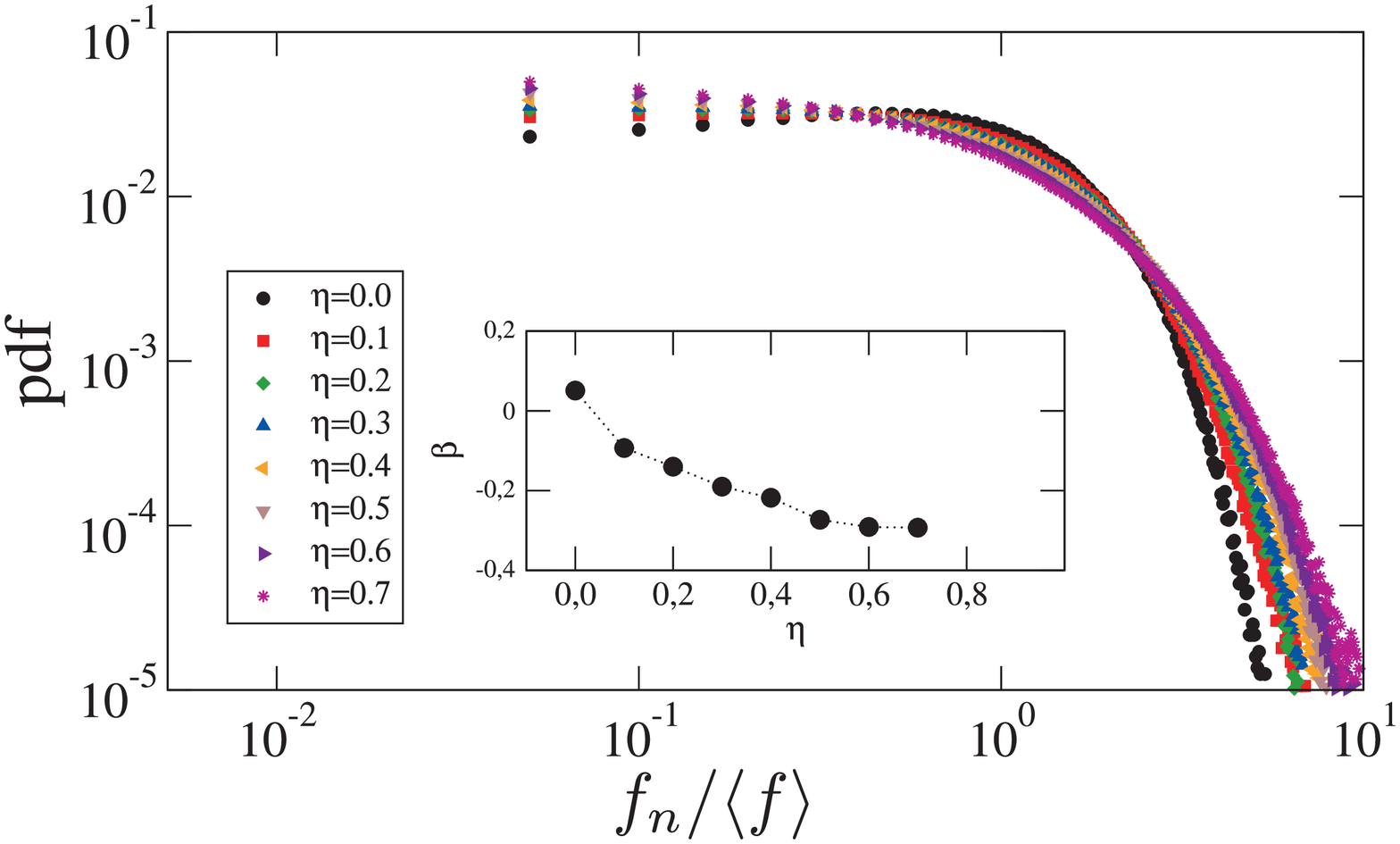} (b)
\caption{Probability distribution function of normal forces $f_{n}$
normalized by the average normal force $\langle f_{n} \rangle$ 
in log-linear (a) and log-log (b) scales for different values of  $\eta$.
\label{sec:weakstrong:pcs_fn}}
\end{figure}

The probability density function (pdf) of normal forces normalized by the mean 
normal force $\langle f_{n} \rangle$
is shown in Fig. \ref{sec:weakstrong:pcs_fn} in log-linear and log-log scales at 
large strains (the data are cumulated from several snapshots in the critical state) for 
all simulated values of $\eta$. As usually observed, in all packings the number of  forces 
above the mean $\langle f_{n} \rangle$ falls off 
exponentially whereas the number of forces below the mean varies as a power-law:
\begin{equation}
P(f_{n}) \propto
\left\{
\begin{array}{lcr}
e^{- \alpha_{n}(\eta)  (f_{n} / \langle f_{n} \rangle)} &, & f_{n}> \langle f_{n} \rangle , \\ 
\left(\frac{f_{n}}{\langle f_{n} \rangle}\right)^{\beta_{n}(\eta)} &, & f_{n}< \langle f_{n} \rangle,
\end{array}
\right.
\label{eqn:strong_fn}
\end{equation}    
where  $\alpha_{n}(\eta)$ and $\beta_{n}(\eta)$ whose variations are 
shown in the insets as a function
of $\eta$. We see that $\alpha_n$ decreases with increasing $\eta$, implying 
that the inhomogeneity of normal forces becomes higher as the particles become 
more elongated. On the other hand,     
$\beta_n$ declines from $0.1$ to $-0.4$ with $\eta$ which means  that the proportion of 
weak contacts (carrying a normal force below the mean) increases with elongation. 
The proportion of weak forces grows from $60\%$ for $\eta=0$  
to $70\%$ for $\eta=0.7$. In other words,
while the proportion of strong contacts declines with increasing $\eta$, 
stronger force chains occur at the same time.

\begin{figure}
\includegraphics[width=8cm]{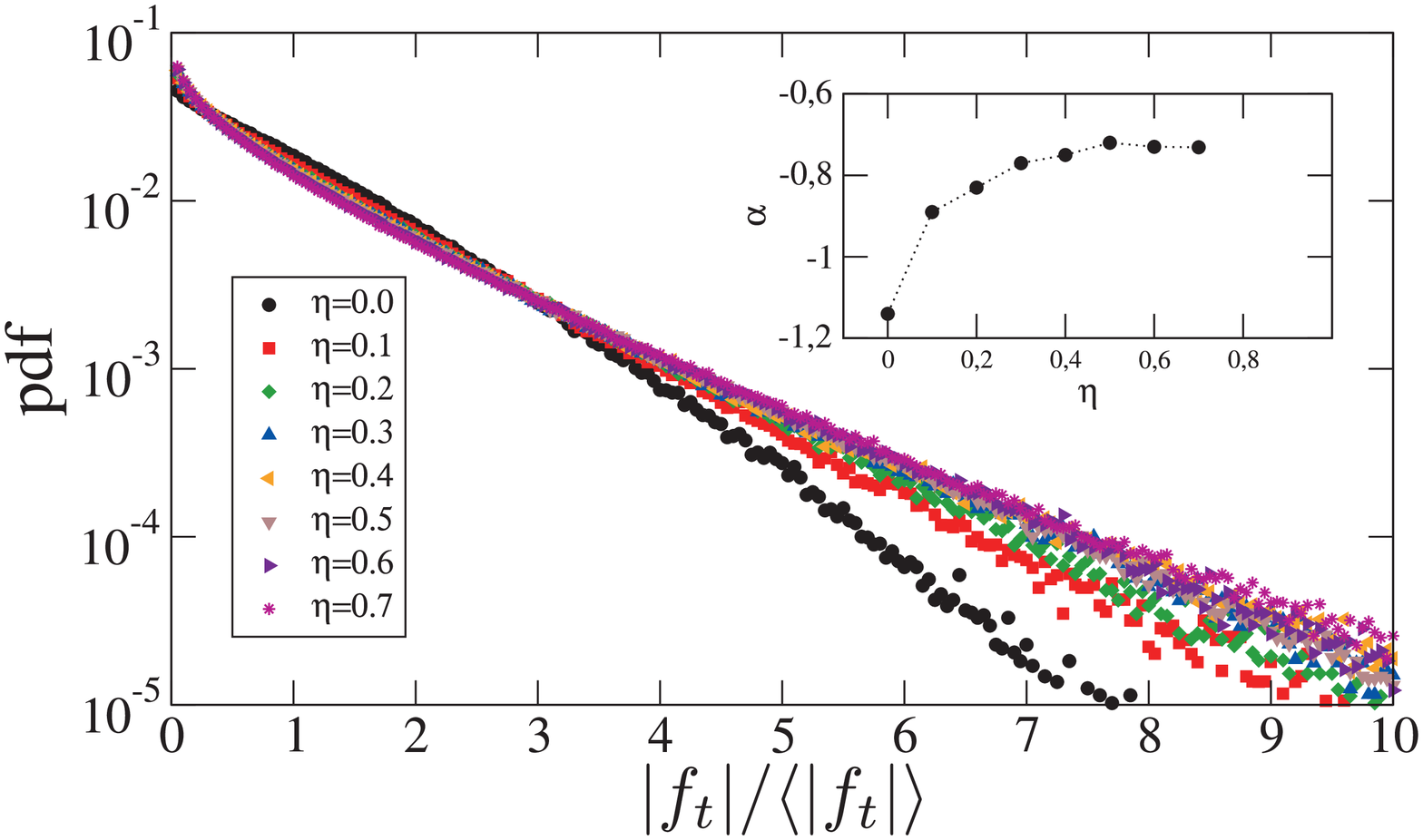}(a)
\includegraphics[width=8cm]{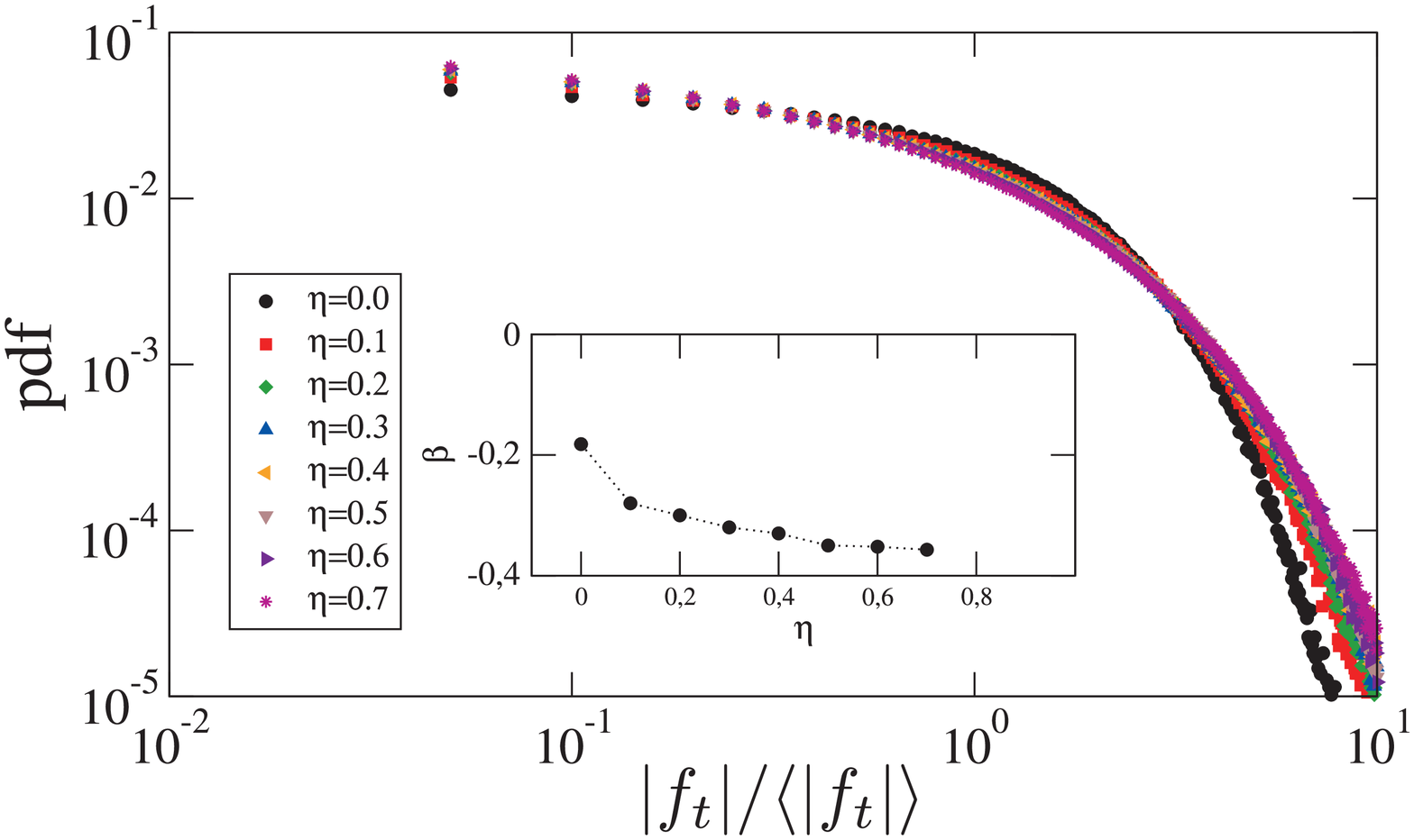}(b)
\caption{Probability distribution function of tangential forces $f_{t}$
normalized by the average tangential force $\langle f_{t} \rangle$ in log-linear (a) 
and log-log (b) scales for different values of  $\eta$.
\label{sec:weakstrong:pcs_ft}}
\end{figure}

Figure \ref{sec:weakstrong:pcs_ft} shows the pdf $P(f_{t})$ 
of tangential forces normalized by the mean tangential 
force $\langle |f_{t}| \rangle$ in each packing. These distributions show also 
an exponential falloff for the forces above the 
average force $\langle |f_{t}| \rangle$ and a 
power law for the forces below $\langle |f_{t}| \rangle$: 
\begin{equation}
P(f_{t}) \propto
\left\{
\begin{array}{lcr}
e^{-\alpha_{t}(\eta)  (|f_{t}| / \langle |f_{t}| \rangle)} &, & |f_{t}|> \langle |f_{t}| \rangle , \\
\left(\frac{|f_{t}|}{\langle |f_{t}| \rangle}\right)^{\beta_{t}(\eta)} &, & |f_{t}|< \langle |f_{t}| \rangle, 
\end{array}
\right.
\label{eqn:strong_ft}
\end{equation}
the corresponding exponents $\alpha_{t}(\eta)$ and $\beta_{t}(\eta)$  
decreasing with $\eta$. We observe that, in contrast to $\alpha_n$ and $\beta_n$, 
the exponents $\alpha_{t}$ and $\beta_t$ saturate beyond $\eta=0.4$. This means 
that the friction forces do not follow the normal forces 
as $\eta$ increases. In other words, the most mobilized (largest) friction forces 
do not occur necessarily at the contacts where the normal forces are 
higher. 

\begin{figure}
\includegraphics[width=8cm]{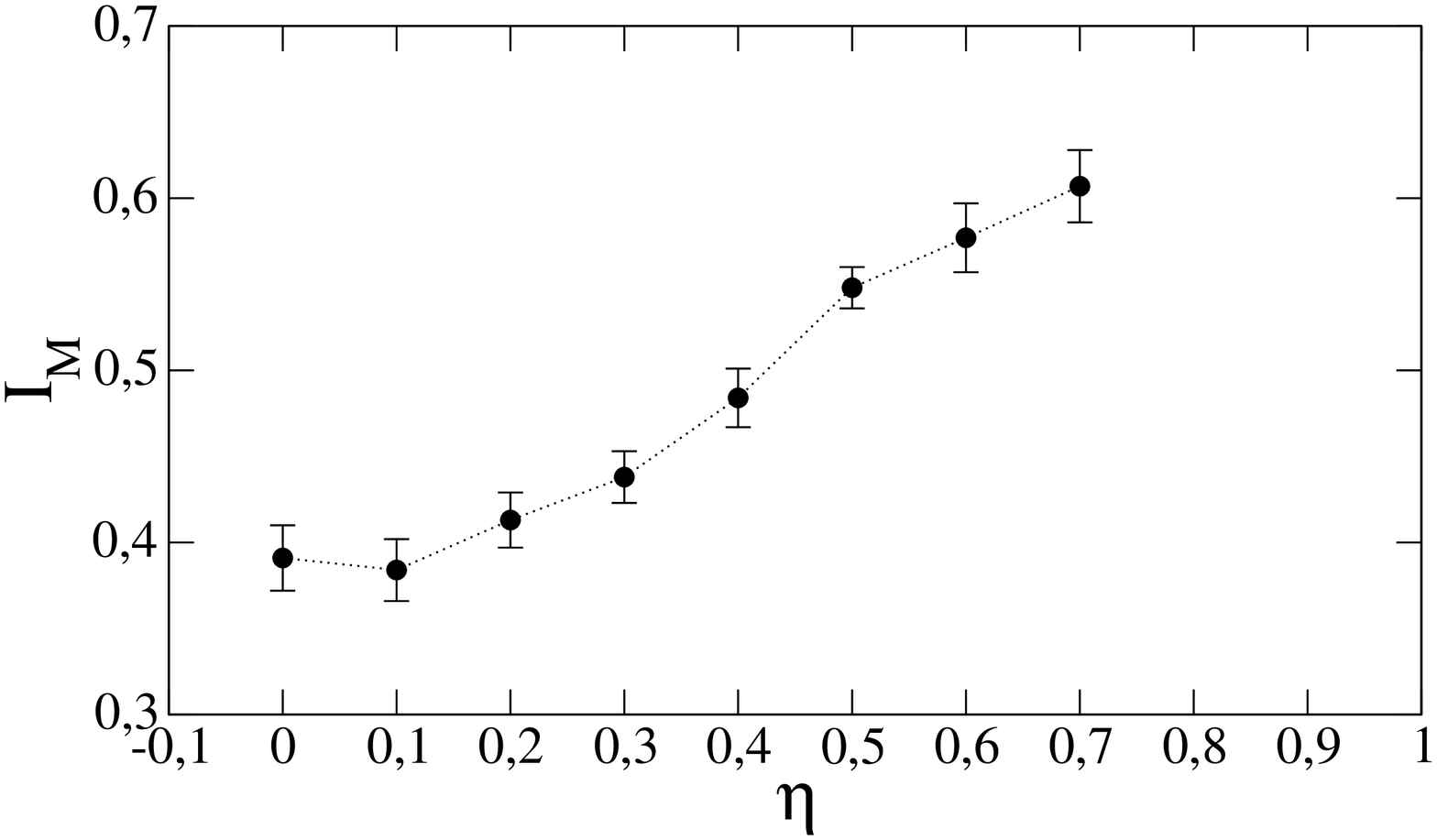}
\caption{Friction mobilization $I_M$ averaged in the steady state as function of  $\eta$.
\label{fig:IM}}
\end{figure}

\begin{figure}
\includegraphics[width=8.5cm]{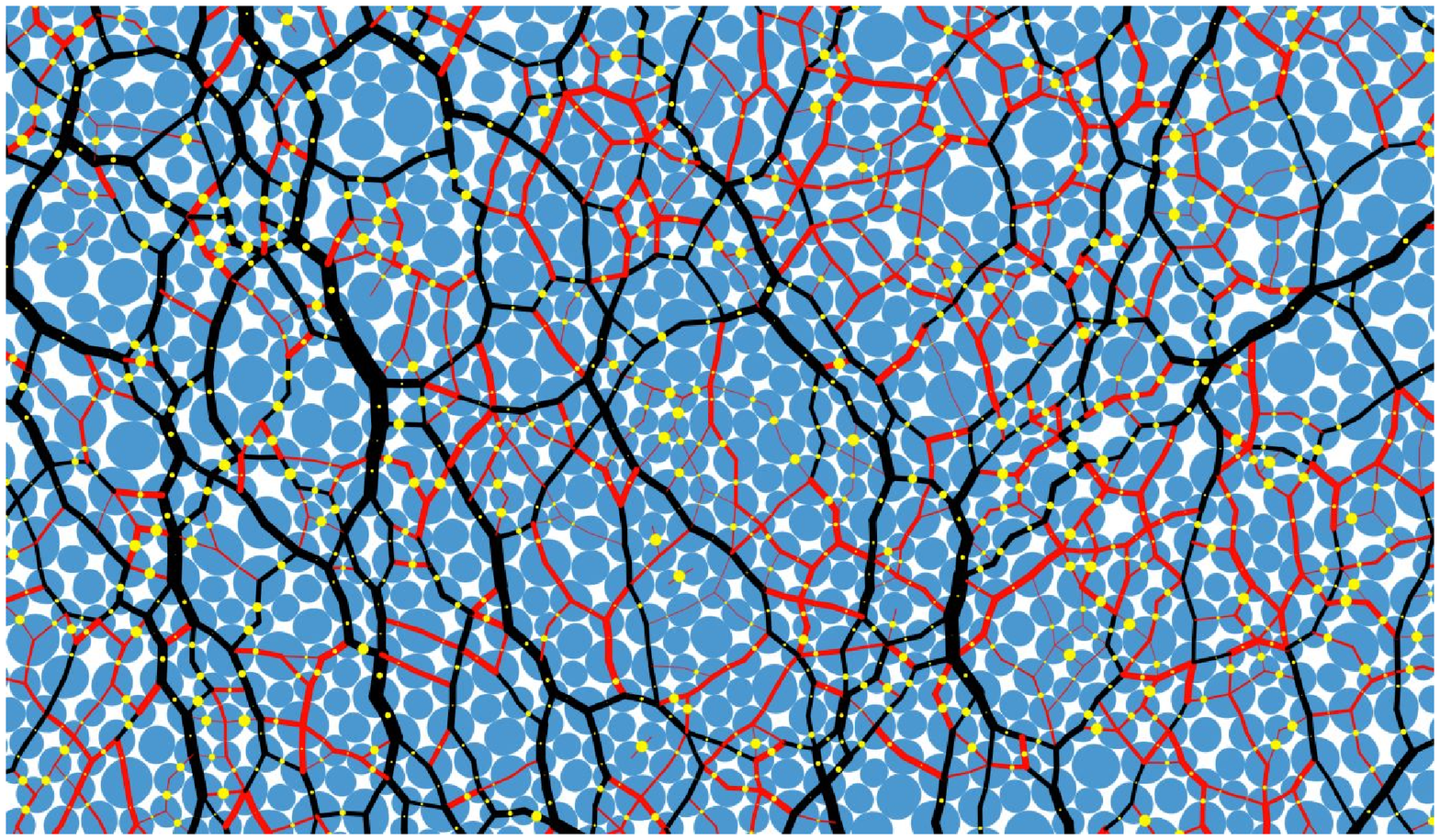}(a)
\includegraphics[width=8.5cm]{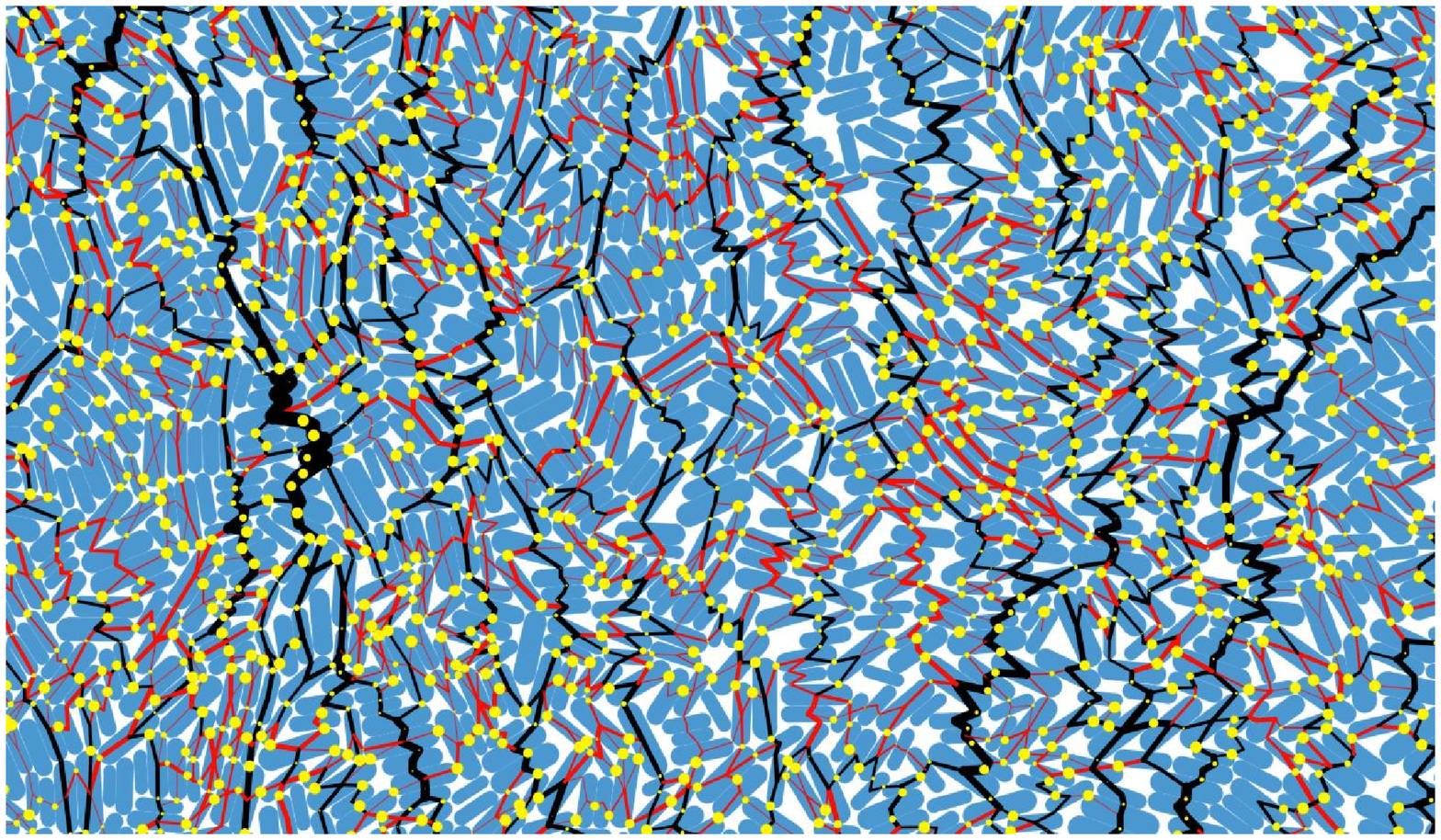}(b)
\caption{A snapshot of the force-bearing particles at $\eta=0.2$(a) and $\eta=0.7$(b) 
and normal forces represented by the thickness of the segments joining the 
particle centers to the application point of the force. 
The strong and weak forces 
are in back and red, respectively. The diameter of yellow circle is proportional
to $I_m$ at the contact.
\label{fig:map_friction_force}}
\end{figure}

\begin{figure}
\includegraphics[width=8cm]{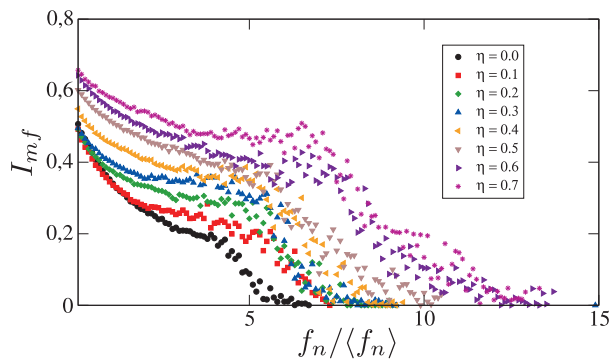}
\caption{Friction mobilization $I_{mf}$  as the average by force class, as a function of $f_n$ for all $\eta$.
\label{fig:IMF}}
\end{figure}

In order to investigate the properties of friction mobilization, we consider the 
friction mobilization index $I_m={|f_t|}/{\mu f_n}$. 
Its average $I_M = \langle \frac{|f_t|}{\mu f_n} \rangle$ 
increases from 0.4 for $\eta=0$ to 0.6 for $\eta=0.7$ as we see in Fig. \ref{fig:IM}. 
This increase underlies to a large extent the increase of the shear strength 
with $\eta$, as we shall see below in Sec. \ref{WS}. However, the friction force is not 
uniformly mobilized  at all contacts. Fig. \ref{fig:map_friction_force} shows a map of weak ($f_n < \langle f_n \rangle$) and strong ($f_n > \langle f_n \rangle$) normal forces, represented by the thickness 
of vectors joining the particle centers to the 
contact points, and the corresponding values of $I_m$,  
represented by circles of diameter proportional to $I_m$ for 
$\eta=0.1$ and$\eta=0.7$. Visual inspection 
reveals that most mobilized contacts belong to the weak force network.   
In fact, the average friction mobilization $I_{mf}$ defined as the 
average by force class, plotted as a function of $f_n$ in Fig. \ref{fig:IMF} 
for all values of $\eta$, declines as $f_n$ increases. We also see that the 
friction mobilization increases with $\eta$ at all force levels.     

\begin{figure}
\includegraphics[width=8cm]{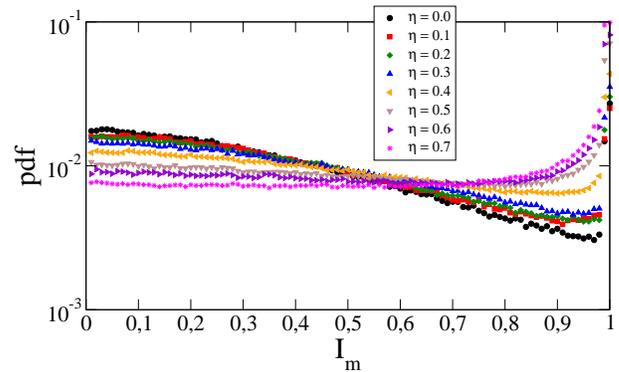}
\caption{Probability distribution function of the friction mobilization index $I_m$.}
\label{fig:pdf_IM}
\end{figure}

Figure \ref{fig:pdf_IM} displays the pdf of  $I_m$  
for different values of $\eta$ in the critical state. For the disks, the pdf is a nearly decreasing linear function 
of $I_m$, which means that the proportion of weakly mobilized  contacts is larger than  
that of strongly mobilized contacts. As $\eta$ is increased, the distribution becomes more 
uniform, and at even larger $\eta$ a class of highly mobilized 
contacts (with $I_m$ close to 1) appears whereas the distribution is nearly uniform for 
all other contacts. This class belongs to weak force network as was shown 
previously, so that not only the friction mobilization $I_m$ 
but also the number of highly mobilized contacts are larger in 
the weak force network. A class of very weak forces was also evidenced 
in \cite{Staron2005a} in a packing of disks deposited under gravity and tilted towards 
its angle of stability. This subclass of the weak network can be defined as the class of  
contacts where the normal force is below the mean but the friction is highly or 
fully mobilized.  

\begin{figure}
\includegraphics[width=9cm]{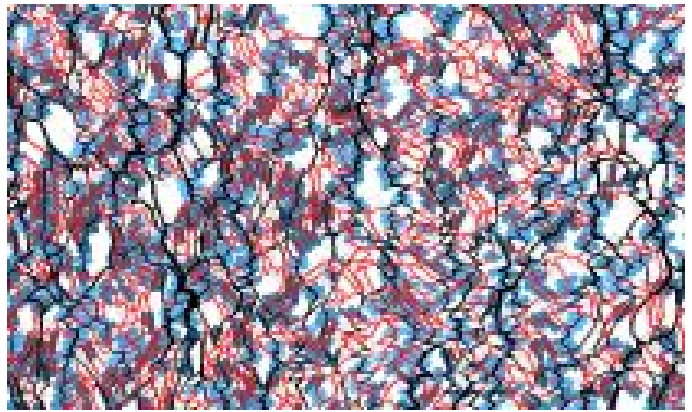}
\caption{A snapshot of the force-bearing particles at $\eta=0.7$ 
and normal forces represented by the thickness of the segments joining the 
particle centers to the application point of the force. The color level for 
the particles is proportional to 
the orientation of the major particle axis for the particles with at least one side/side 
contact. The particles having no side/side contacts are in gray. The strong and weak forces 
are in back and red, respectively.}
\label{fig:nematic}
\end{figure}

This enhanced friction mobilization implies that 
the equilibration of the particles is more complex than in disk packings. 
In particular, the nematic ordering due to the ``geometrical'' chains 
of side/side contacts between particles means that the statistics of 
forces and the mobilization of friction are closely related to 
the equilibrium of such chains rather than that of individual particles. 
These chains are evidenced in Fig. \ref{fig:nematic} for $\eta=0.7$ 
where the force bearing 
particles belonging to the chains are represented by a color 
level proportionally to their orientations. The friction needs to be 
highly mobilized  inside the chains in order to ensure their stability.            

\subsection{Branch vectors}

\begin{figure}
\includegraphics[width=3cm]{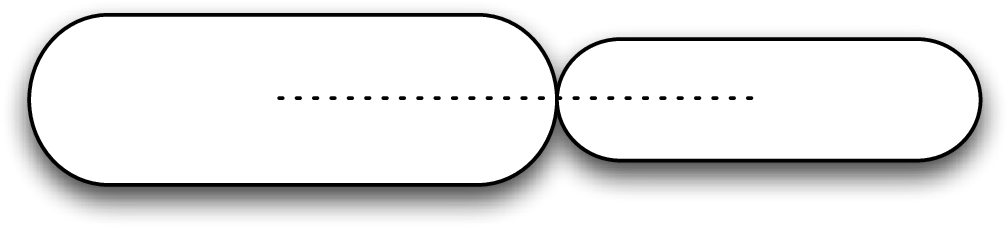}(a)
\includegraphics[width=2cm]{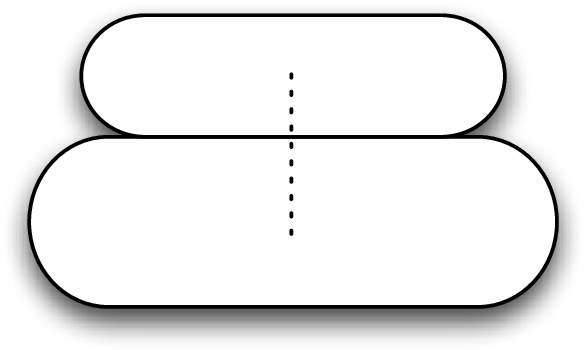}(b)
\includegraphics[width=2cm]{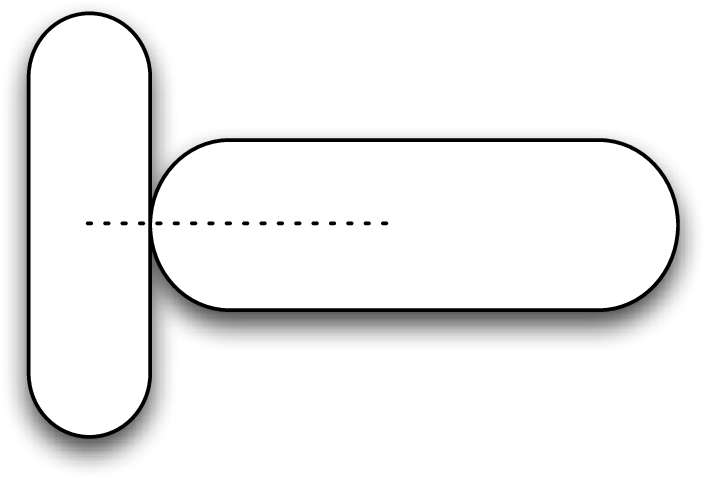}(c)
\caption{Principals modes of contacts : cap-cap (a), side-side (b) and cap-side (c).
\label{fig:sketch_contact}}
\end{figure}

The branch vectors in a packing of elongated particles reflect both the 
relative orientations of the particles in contact and their size distribution. 
The latter may be integrated out by simply dividing the branch vector length $\ell$ 
between two touching particles by the sum $R_1+R_2$ of the radii  
$R_1$ and $R_2$   of their circumscribing circles. This {\em reduced}  
branch-vector length $\ell_r = \ell /(R_1+R_2)$ varies in the range $[1-\eta, 1]$.  
We have $\ell_r=1$ at $\eta=0$ (for disks). 
For elongated particles, $\ell_r=1$ corresponds to a cap/cap contact between 
two aligned particles, Fig. \ref{fig:sketch_contact}(a), whereas $\ell_r=1-\eta$ corresponds to 
a centered side/side contact between two parallel particles, Fig. \ref{fig:sketch_contact}(b). 
Such contact configurations, when they exist,  can be evidenced 
from the probability density function of $\ell_r$ and its possible modes  
at $\ell_r=1$ or $\ell_r=1-\eta$. 

\begin{figure}
\includegraphics[width=9cm]{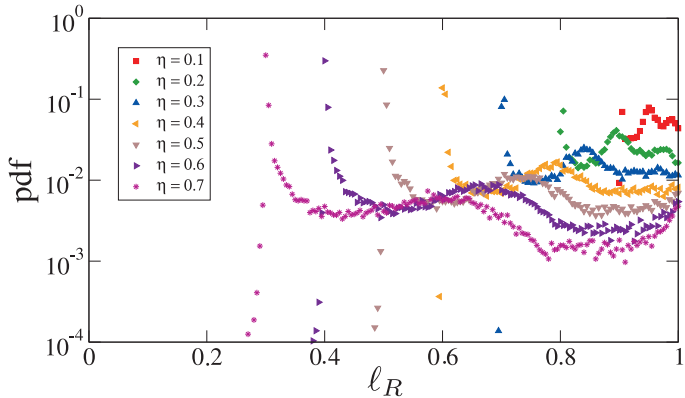}
\caption{Probability distribution function of the reduced 
branch-vector lengths $\ell_r$ for all values of $\eta$ in the critical state.
\label{fig:pdf_lr}}
\end{figure}

\begin{figure}
\includegraphics[width=10cm]{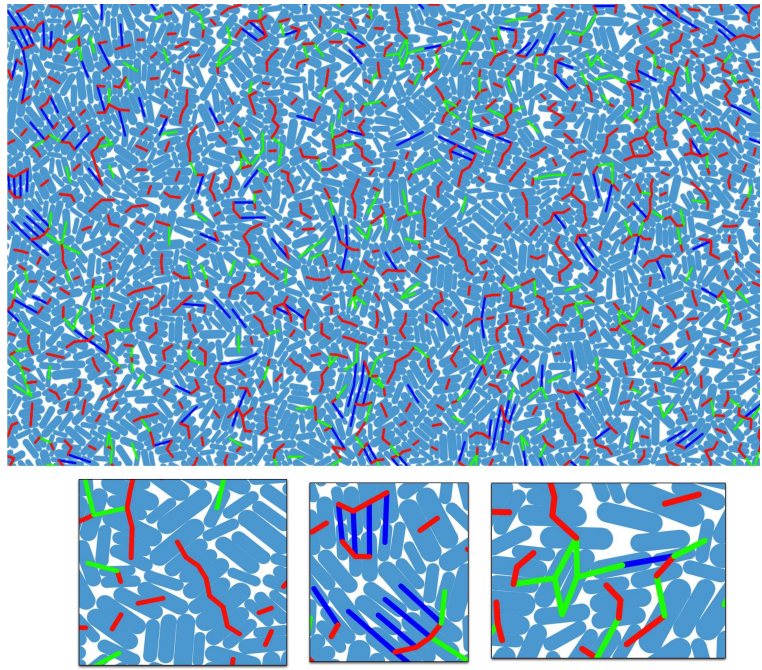}
\caption{A snapshot of cap-cap modes contact (blue), side-side modes contacts (red) and cap-side modes contacts (green).
\label{fig:Map_type_contact}}
\end{figure}

Figure \ref{fig:pdf_lr} displays the pdf of reduced 
branch-vector lengths for all values of $\eta$ in the critical state. 
These pdf's are nothing but normalized radial functions with 
$\ell_r$ varying in a limited range as only the touching particles are considered. 
They are nearly similar for all values of $\eta$. 
The first mode, centered on $\ell_r=1-\eta$ reveals the presence of 
a broad population of side/side contacts with a peak increasing in amplitude 
with $\eta$ as displayed in Fig. \ref{fig:Map_type_contact} (b). 
We also observe a less pronounced mode, centered on  $\ell_r \simeq 1$, 
corresponding to a distinct population of aligned cap/cap contacts, also marked in 
Fig. \ref{fig:Map_type_contact} (a). The intermediate mode occurs approximately at  
$\ell_r \simeq 1-\eta/2$ which is the midpoint of  
the interval $[1-\eta, 1]$. This length corresponds to an orthogonal 
side/cap contact as shown in Fig. \ref{fig:sketch_contact}(c). The presence of such a distinct 
mode, through decreasing in amplitude as $\eta$ increases, is a clear proof of the 
occurrence of orthogonal layers some of which are observed 
in Fig. \ref{fig:Map_type_contact}. This mode is also characterized by a broad extension 
reflecting the intermediate angles between the orientations of touching 
particles.     

\begin{figure}
\includegraphics[width=8cm]{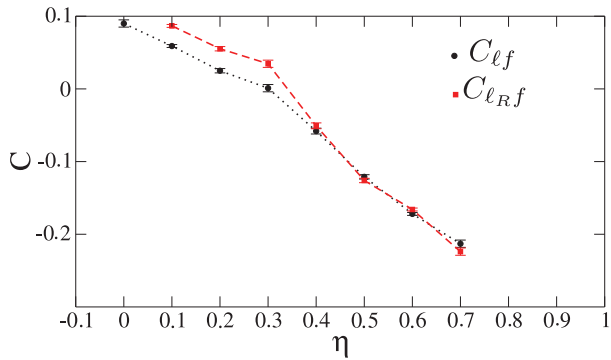}
\caption{Correlation $C_{\ell_R f}$ and $C_{\ell f}$ as a function of $\eta$ averaged in the critical state.
\label{fig:correlation_flr}}
\end{figure}

We expect the branch vectors lengths to be correlated with contact forces 
because of either the contact configurations they represent or simply 
the fact that force chains tend to be captured by larger particles (hence, longer 
branch vectors) \cite{Voivret2009}. 
This correlation can be estimated with the Pearson coefficient, which for 
two random variables $x$ and $y$ is defined by the scalar 
\begin{equation}
C_{xy} = \frac{\langle (x - \langle x \rangle)(y - \langle y \rangle) \rangle}{\sqrt{\langle (x - \langle x \rangle)^2 \rangle}{\sqrt{\langle (y - \langle y \rangle)^2 \rangle}}},
\label{eq:pearson}
\end{equation}
Note that $C=1$ corresponds 
to a full inter-dependence whereas $C=0$ means   
full statistical independence of the two variables. 
Fig. \ref{fig:correlation_flr} shows the Pearson coefficients   
$C_{f\ell_r}$, between the force amplitude $f$ and $\ell_r$, as 
well as $C_{f\ell}$, between $f$ and $\ell$, as a function of $\eta$. 
Both coefficients decrease with $\eta$ from positive values for $\eta \leq  0.3$ to 
negative values down to $-0.22$. The positive correlation (larger forces at longer 
branch vectors) is a consequence of the fact that the distribution of 
branch lengths at low values of $\eta$ is governed by particles sizes. On the other hand, 
the negative correlation (larger forces at shorter branch vectors) reflects the 
effect of the increasing number of side/side contacts as the particles 
become more elongated.

\begin{figure}
\includegraphics[width=8cm]{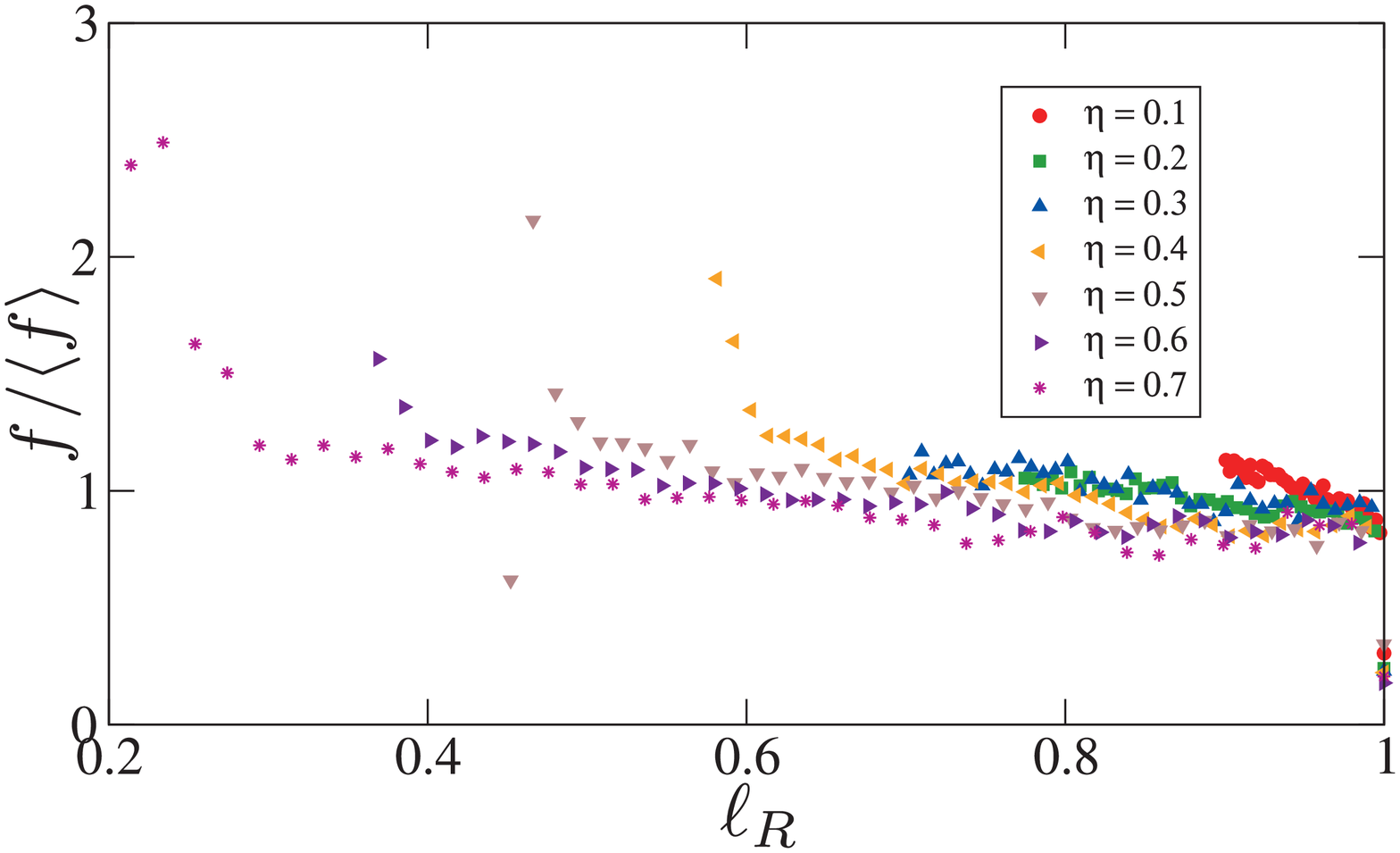}(a)
\includegraphics[width=8cm]{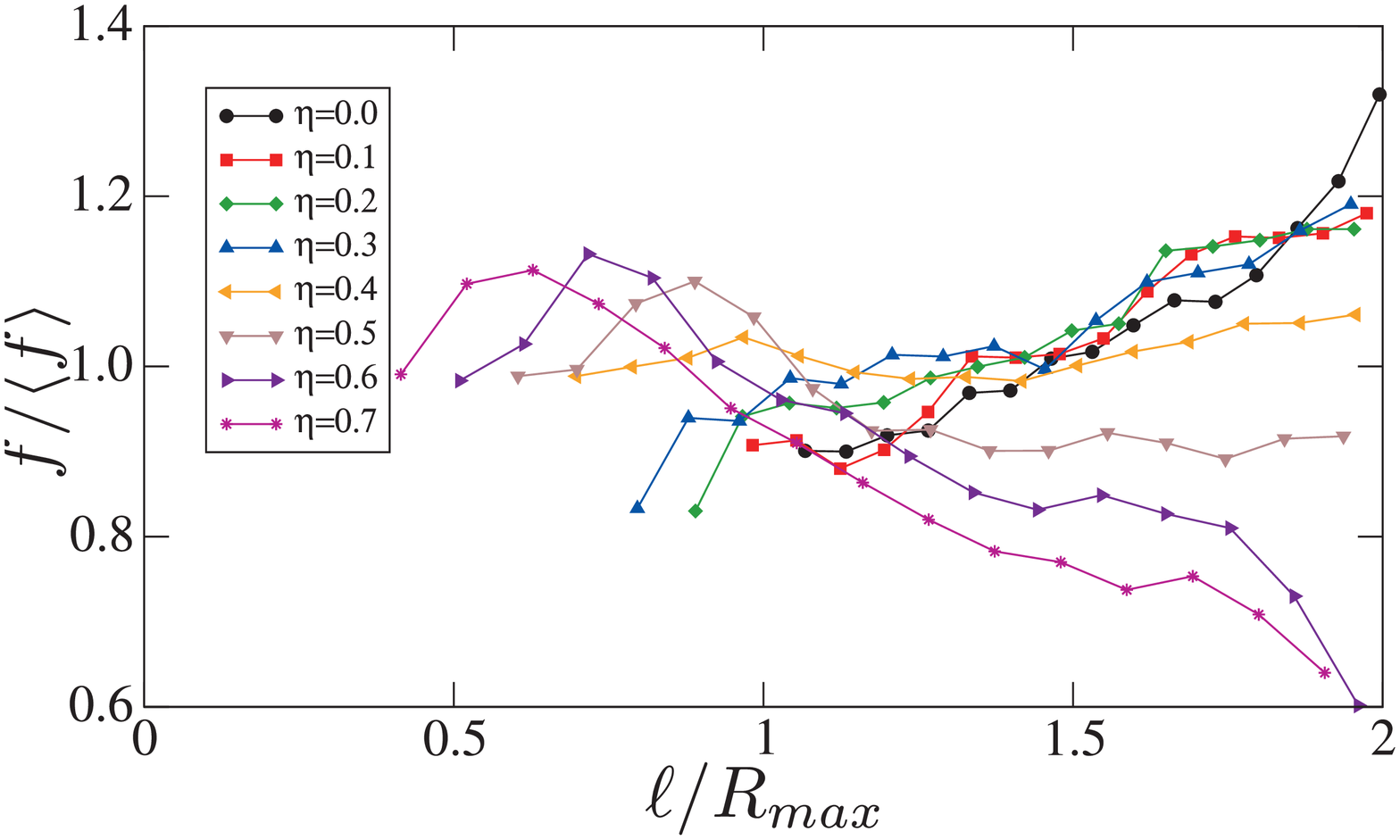}(b)
\caption{Linear correlation between contact force $f$ and branch length $\ell$ as a function of $\eta$.
\label{fig:Flr_Correlation}}
\end{figure}

Further insight into this force/branch-length correlation can be obtained 
from the average force amplitude $\langle f \rangle_{\ell_r}$, calculated 
by taking the average force in a class of contacts in the interval 
$[\ell_r - \Delta \ell_r/2,\ell_r + \Delta \ell_r/2]$,   
as a function of $\ell_r$, shown in Fig. \ref{fig:Flr_Correlation}(a)  
for all values of $\eta$. 
This plot shows that for all contact classes the associated mean force $\langle f \rangle_{\ell_r}$  
is nearly equal to the global mean force $\langle f \rangle$ except to the 
class of the shortest branch vectors (side/side mode), which concentrates a 
mean force above  $\langle f \rangle$, and the class of the longest 
branch vectors (cap/cap mode), which seem to carry a considerably 
lower force on the average. In his way, the rather weak correlation between the reduced 
branch length and force appears here to be governed by the two afore-mentioned modes. 
In order to evidence the effect of particle size distribution, let us consider 
the average force amplitude $\langle f \rangle_{\ell}$ 
as a function of $\ell$ as shown in Fig. \ref{fig:Flr_Correlation}(b)  
for all values of $\eta$. For $\eta \leq 0.3$, the contact force is on the average 
an increasing function of $\ell$. For disks ($\eta=0$), the variation of $\ell$ 
is a consequence only of the particle size distribution and, therefore, the increase of 
the mean force with $\ell$ means that the larger particles, involved in the 
longer branch vectors, capture higher forces. The same effect  
seems thus to underly also the increasing mean force with $\eta$ for   
elongated particles with $\eta \leq 0.3$. But at larger elongations, 
the trend is reversed and we see that the mean force declines as $\ell$ 
increases, reflecting thus the effect of the side/side contact mode as discussed 
previously.

\section{Weak and strong force networks}
\label{WS}

The complex network of contact forces in a packing of elongated particles 
can also be analyzed by considering the contribution of various classes of 
forces and/or branch vectors to stress transmission. Indeed, according to 
equation (\ref{eq:sigma}), the stress tensor is expressed as an average involving 
branch vectors and contact forces, so that partial summations allow one to 
define partial stress tensors that have been applied in the past 
to investigate the scale-up of local quanties \cite{Radjai1998}. For example, the subset of 
contacts carrying a force below a  threshold, reveals the 
respective roles of weak and strong force chains with respect to the overall shear strength 
of granular materials \cite{Radjai1998}. In this section, we 
apply this methodology to analyze the stress and other texture-dependent quantities 
in view of elucidating the effect of particle elongation. 

In what follows, we consider various fabric and force parameters for 
the  ``$\xi$-networks'' defined as the subsets $\mathcal{S}(\xi)$ of contacts which carry
a force below a cutoff force $\xi$ normalized by the 
mean force (ie $f/\langle f_n \rangle \in [0,\xi]$), where
$\xi$ is varied from $0$ to the maximal force in the system. 
The {\em weak} network corresponds to $\mathcal{S}(1)$ whereas the
{\em strong} network is its complement. 
In section \ref{sec:distribution}, we focused on scalar descriptors of 
granular texture  such as the distributions and correlations of 
force magnitudes and branch lengths. Beyond these low-order 
quantities, the granular texture is characterized by a disordered but anisotropic structure 
of both the contact and force  networks, which require higher-order description 
in terms of various fabric and force tensors. 
We analyze below different 
parameters pertaining to this tensorial organization of our packings 
as a function of $\xi$ and for increasing elongation $\eta$.

\subsection{Granular texture}

A relevant description of granular texture is given 
by  the probability distribution $P(\bm n)$
of the contact normals $\bm n$ ; see Fig. \ref{sec:numerical_procedure:frame}.
In two dimensions, the unit vector $\bm n$ is described by a single angle $\theta \in [0,\pi]$.
The distribution $P_\theta(\theta)$ of contact orientations can be evaluated from  
the numerical data at different stages of its evolution. 
In our simulations, all numerical samples are prepared 
in an isotropic state so that $P_\theta = 1/\pi$ in the initial state.  
This distribution evolves with shear strain and becomes  increasingly 
more anisotropic as the critical state is approached. 
By restricting the data to those belonging to the 
$\xi$-networks, we obtain a continuous family of distributions $P_\theta (\theta,\xi)$  
that describe the geometrical state of the system. In practice, however, 
such functions can be estimated with meaningful statistics only in 
the critical  state where the data can be cumulated from independent 
configurations representing all the same state. 

\begin{figure}
\includegraphics[width=6cm]{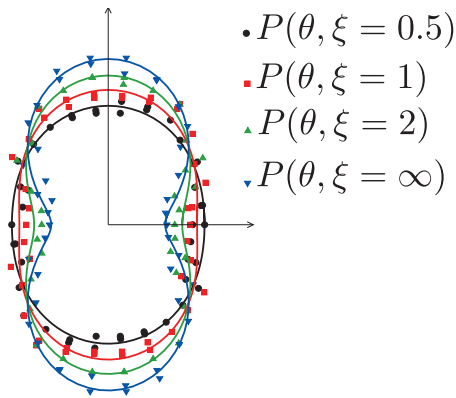}
\caption{Distributions of contact orientations (symbols) in polar 
coordinates for $\eta=0.5$ and 
several values of the force cutoff $\xi$ together with their 
Fourier fits (\ref{eqn:harmonique}) (full lines).}
\label{fig:Ptheta}
\end{figure}

Figure \ref{fig:Ptheta} shows  the distributions $P_\theta (\theta,\xi)$ 
in polar coordinates for $\eta=0.5$ and for several values of $\xi$. 
The distributions are similar with nearly the same privileged 
direction aligned with the principal stress direction $\theta_\sigma$ 
but with increasing anisotropy as a function of $\xi$. They all can  be 
approximated by their truncated Fourier expansion \cite{Radjai1998,Azema2007,Estrada2008}:
\begin{equation}
\label{P_theta_xi}
\begin{array}{lcl}
P_\theta (\theta,\xi) &=& \frac{1}{2\pi}  \{ 1 + a_c(\xi) \cos 2(\theta - \theta_\sigma) \},  \\
\end{array}
\end{equation}
where $a_c(\xi) $ is the amplitude of contact anisotropy in the $\xi$-network. 
In practice, it is more convenient to estimate $a_c(\xi) $  through the  partial fabric 
tensors $\bm F (\xi)$ defined by \cite{Satake1982}:
\begin{equation}
F_{\alpha \beta}(\xi) = 
\frac{1}{\pi} \int_0^\pi  n_\alpha(\theta, \xi) n_\beta (\theta, \xi)  P_\theta(\theta,\xi) d\theta, 
\label{eq:F_xi}
\end{equation}
where $\alpha$ and  $\beta$ design the cartesian components.   
By definition, we have $tr ({\bm F}(\xi)) = 1$. Introducing the harmonic  expression 
(\ref{P_theta_xi}) in (\ref{eq:F_xi}), we get 
\begin{equation}
a_c(\xi) = 2(F_{1}(\xi)-F_{2}(\xi)) \cos 2[\theta_c(\xi) - \theta_\sigma], 
\label{eqn:harmonique}
\end{equation}
where the subscripts $1$ and $2$ refer to the principal 
values of $\bm F (\xi)$ and $\theta_c(\xi)$ represents the privileged direction 
of the  partial fabric 
tensors $\bm F (\xi)$. Note that, up to statistical fluctuations, 
the principal directions of the fabric and stress tensors coincide 
in the critical state for each $\xi$-network, so that the 
phase factor $\cos 2[\theta_c(\xi) - \theta_\sigma]$ is 
either equal to 1  when $\theta_c(\xi) = \theta_\sigma$ or equal to $-1$ when 
$\theta(\xi) = \theta_\sigma + \pi/2$. 

\begin{figure}
\includegraphics[width=8cm]{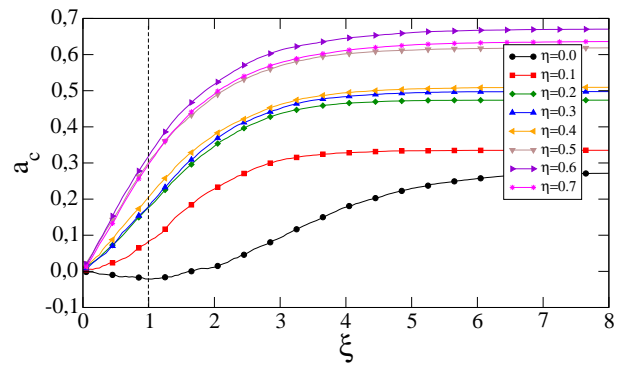}
\caption{Partial fabric anisotropy $a_c$ as a function of force cutoff $\xi$ 
normalized by the mean force $\langle f \rangle$ for different values of  $\eta$.
\label{sec:fabric_force_transmission:ac}}
\end{figure}

Fig.\ref{sec:fabric_force_transmission:ac} displays $a_c$ as a function 
of $\xi$ for all values of $\eta$. For the disk packings ($\eta=0$),   
the anisotropy of weak contacts is negative but increases in absolute value and  
reaches its peak value at $\xi\sim1$. This negative value indicates that in disk packings 
the weak contacts are orientated preferentially perpendicular to 
the major principal stress direction \cite{Radjai1998}.  
As more contacts come into play with increasing $\xi$, the partial 
anisotropy $a_c(\xi)$ becomes less negative and finally changes sign, showing that the 
strong contacts are mainly along the major principal stress direction. 
This bimodal behavior of stress transmission is a nontrivial organization of the force network 
and holds also in 3D in the case of sphere packings \cite{Azema2009}.   
However, it is remarkable that for elongated particles ($\eta>0$),  
the partial anisotropies of both weak and strong networks are positive, as observed in 
Fig. \ref{sec:fabric_force_transmission:ac}. This means that, in contrast to 
the disk packings, the weak and strong contacts 
in  packings of elongated particles can not be differentiated on the basis of their 
roles in the $\xi$-networks. Physically, this behavior may be interpreted 
by stating that the 
static equilibrium of the chains of elongated particles does not require the stabilizing effect 
of the weak contacts. 
A similar result was observed by Estrada et al. for disk packings 
at large values of rolling resistance, which allows for the equilibrium of 
long chains of particles inter-connected by only two contacts \cite{Estrada2008}. 
But, as we shall see below, for our elongated particles the 
differentiation between the two networks operates via 
the forces carried by the $\xi$-networks.  

\begin{figure}
\includegraphics[width=9cm]{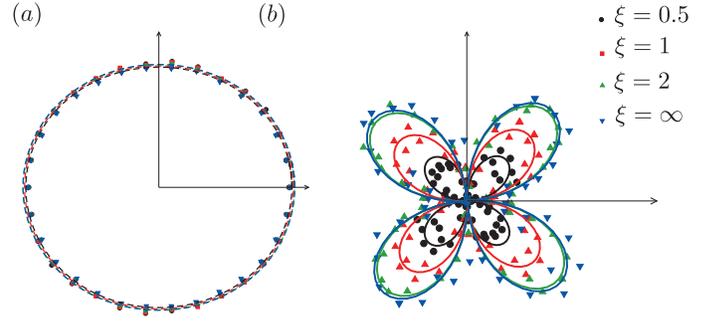}
\caption{Distributions of $\langle \ell_{n} \rangle (\theta,\xi) $(a) and $\langle \ell_{t} \rangle (\theta,\xi)$(b) (symbols) in polar 
coordinates for $\eta=0.5$ and 
several values of the force cutoff $\xi$ together with their 
Fourier fits (\ref{Aniso_values_xi}) (full lines).}
\label{fig:Distribution_lnlt}
\end{figure}

The information involved in the angular distribution $P_\theta$ may  
be enriched by accounting for the branch vectors $\bm \ell$ which, 
as seen in section \ref{sec:distribution}, reflects both the particle size 
distribution and local contact modes. We thus consider here the      
average normal and tangential branch 
vector components $\langle \ell_n \rangle (\theta,\xi)$ and 
$\langle \ell_t \rangle (\theta,\xi) $ defined in (\ref{eq:n}), obtained by 
averaging $\ell_n$ and $\ell_t$ over the contacts oriented along $\theta$ within a 
centered angular interval $\Delta \theta$.  As for $P_\theta$, we evaluate 
these functions in the critical state, for different values of $\eta$ and 
as $\xi$. Fig. \ref{fig:Distribution_lnlt} shows the functions $\langle \ell_n \rangle (\theta,\xi) $ 
and $\langle \ell_t \rangle (\theta,\xi) $ in polar coordinates for $\eta=0.5$ 
and for several values of $\xi$. 
These functions are anisotropic with an anisotropy which depends on 
$\xi$. We introduce here their truncated expansion on an orthonormal 
Fourier basis:
\begin{equation}
\label{ell_theta_xi}
\left\{
\begin{array}{lcl}
\langle \ell_{n} \rangle (\theta,\xi) &=&  \langle \ell \rangle    \{ 1 + a_{ln}(\xi)\cos 2(\theta - \theta_{\sigma}) \}, \\
\langle \ell_{t} \rangle (\theta,\xi) &=&  \langle \ell \rangle  a_{lt}(\xi) \sin 2(\theta - \theta_{\sigma}),
\end{array}
\right.
\end{equation}
where $a_{ln}(\xi)$ and $a_{lt}(\xi) $ are the normal and tangential branch 
anisotropies in the $\xi$-networks. Note that, by construction  we have  
$a_{lt}=0$ for disks ($\eta=0$).
The analytical form of $\langle \ell_t \rangle (\theta,\xi) $ results from the orthonormal nature 
of the Fourier basis  and the 
fact that the mean value of $\ell_t$ vanishes due to disorder:
\begin{equation}
\int_0^\pi \langle \ell_t \rangle (\theta,\xi) \ P_\theta (\theta,\xi) \ d\theta =0. 
\end{equation}
Fig. \ref{fig:Distribution_lnlt} shows that this functional form provides a 
good approximation of the data.    

For the calculation of $a_{ln}(\xi)$ and $a_{lt}(\xi)$, we introduce the following 
{\it branch tensors} \cite{Azema2010}: 
\begin{equation}
\label{Chi_tensors_xi}
\left\{
\begin{array}{lcl}
\chi^{ln}_{\alpha \beta}(\xi) &=& 
\frac{1}{\langle \ell \rangle} \int\limits_{0}^\pi  \langle \ell_{n} \rangle(\theta, \xi)  n_\alpha(\xi)  n_\beta(\xi) P_ \theta(\theta, \xi) d \theta,  \\ 
\chi^{lt}_{\alpha \beta}(\xi) &=& 
\frac{1}{\langle \ell \rangle} \int\limits_{0}^\pi  \langle \ell_{t} \rangle(\theta, \xi)  n_\alpha(\xi)  t_\beta(\xi) P_ \theta(\theta, \xi) d \theta,  
\end{array}
\right.
\end{equation}      
The following relations are then easily obtained:
\begin{equation}
\label{Aniso_values_xi}
\left\{
\begin{array}{lcl}
a_{ln}(\xi) &=& 2[\chi^{ln}_{1}(\xi) - \chi^{ln}_{2}(\xi)]  /   {\mbox tr}[ \bm \chi^{ln}(\infty)] - a_c(\xi),  \\ 
a_{lt}(\xi)&=& 2[\chi^{l}_{1}(\xi) - \chi^{l}_{2}(\xi)]        /   {\mbox tr}[ \bm \chi^{l}(\infty)]          - a_c(\xi) - a_{ln}(\xi),  
\end{array}
\right.
\end{equation}      
where  $\bm \chi^{l} = \bm \chi^{ln} + \bm \chi^{lt}$, and
the subscripts $1$ and $2$ refer to the principal values of each tensor. 
By construction, we have 
$tr \bm \chi^{l} = (\chi^{l}_{1} + \chi^{l}_{2})=\langle \ell \rangle$.  Note also that 
the two partial branch vector anisotropies $a_{ln}$ and $a_{lt}$ may take positive or 
negative values depending on the orientations $\theta_{ln}$ and $\theta_{lt}$ 
of the two tensors with respect to $\theta_\sigma$.   

\begin{figure}
\includegraphics[width=8cm]{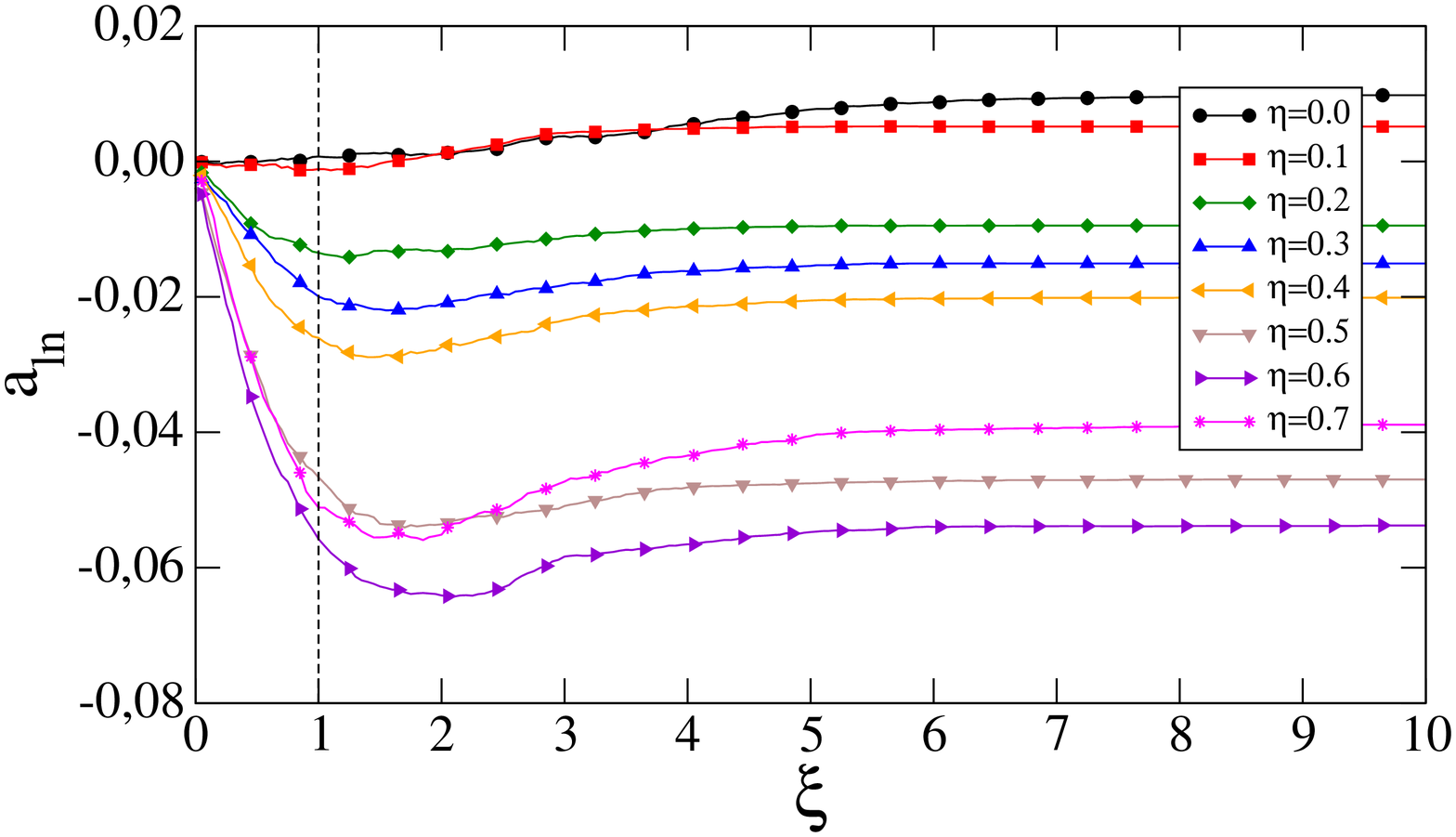}
\includegraphics[width=8cm]{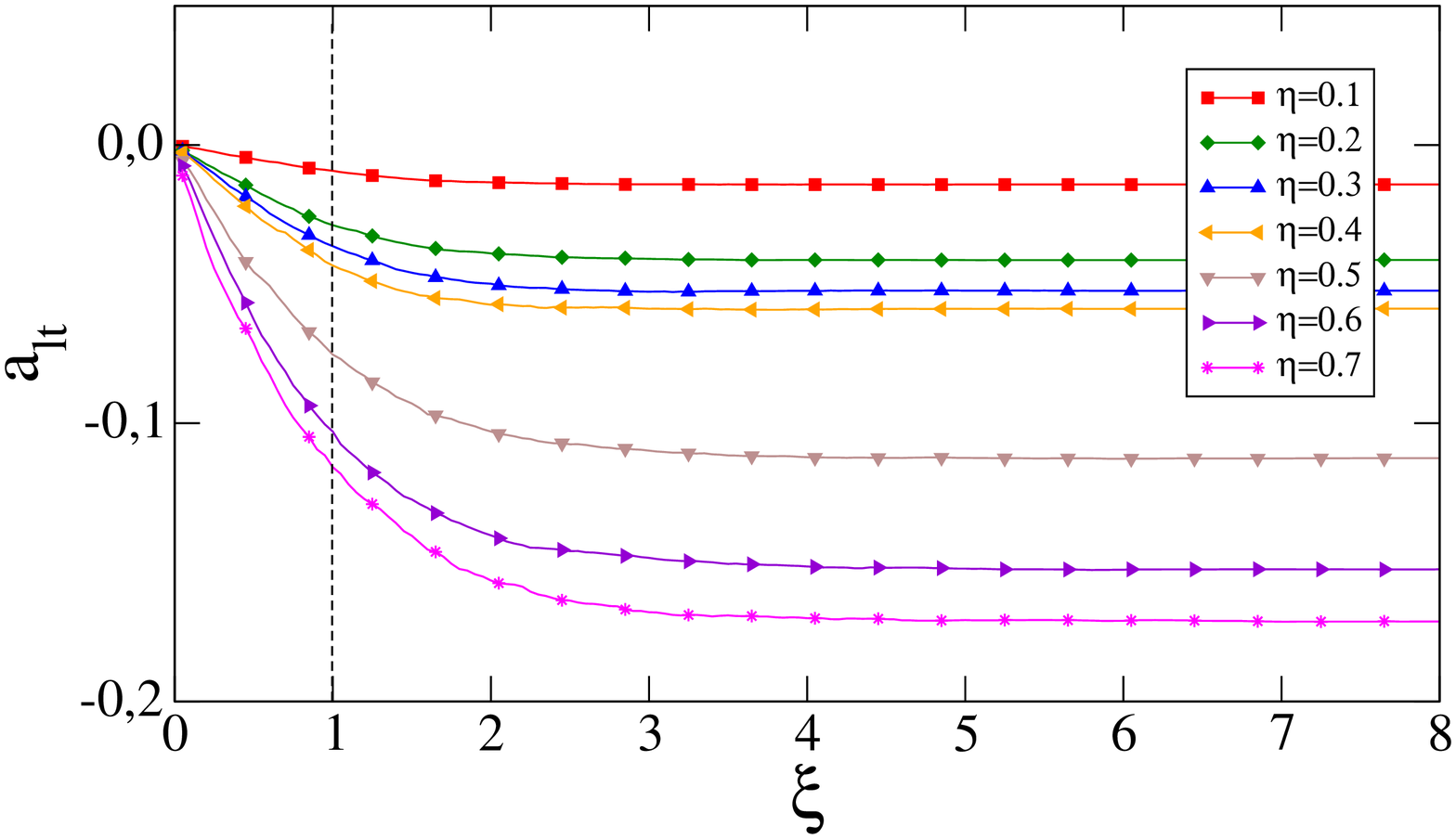}
\caption{Partial normal and tangential branch vector length anisotropies $a_{ln}$ and 
$a_{lt}$   as a function of force cutoff $\xi$ normalized by the mean 
force $\langle f \rangle$ for different values of  $\eta$.
\label{sec:fabric_force_transmission:alnt}}
\end{figure}

Figure \ref{sec:fabric_force_transmission:alnt} shows 
the branch-vector anisotropies $a_{ln}(\xi)$ and $a_{lt}(\xi)$ 
as a function of $\xi$ in the critical state for all values of $\eta$. 
$a_{ln}(\xi)$ is positive for $\eta=0$ and $\eta=0.1$ and increases slightly with 
$\xi$, but for more elongated particles it takes negative values, which means 
that the particles tend to form longer 
branch vectors with their neighbors in the direction of extension. 
As $\xi$ increases, this anisotropy increases in absolute value and 
reaches a plateau after passing by a peak value at a point in the range 
$\xi \in [1,2]$. This behavior suggests 
that the particles touch preferentially along their minor axes when 
the contact orientation is close to 
the compression axis (in the strong network), and along their major axis 
when the contact orientation is close to 
the extension axis (in the weak network), in agreement with the fact that 
the longest branches are in the weak network ; see Sec. \ref{sec:distribution}.
As to $a_{lt}(\xi)$, its value is always negative and  increases monotonically 
with $\xi$ in absolute value. 
Note also that, for all values of $\xi$,   
$a_{lt}(\xi)$ is much higher than $a_{ln}(\xi)$ while both remain weak compared to 
$a_c(\xi)$. 

\subsection{Force anisotropies}

\begin{figure}
\includegraphics[width=9cm]{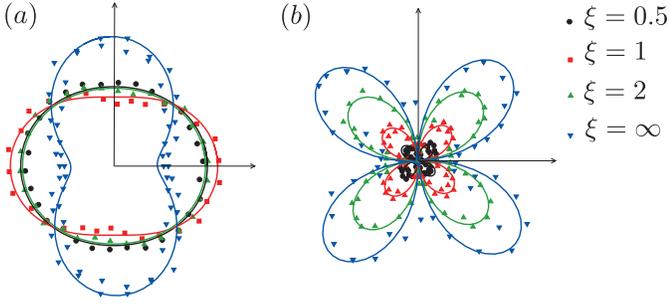}
\caption{Distributions of $\langle f_{n} \rangle (\theta,\xi) $(a) and $\langle f_{t} \rangle (\theta,\xi)$(b) (symbols) in polar 
coordinates for $\eta=0.5$ and 
several values of the force cutoff $\xi$ together with their 
Fourier fits (\ref{ell_f_theta_xi}) (full lines).}
\label{fig:Distribution_fnft}
\end{figure}

We now consider the angle-averaged normal and tangential forces, $\langle f_n \rangle(\theta,\xi) $ 
and $\langle f_t \rangle(\theta,\xi) $ in the $\xi$-network.
A second order Fourier expansion provides an adequate representation 
of these distributions for all values of $\xi$ as shown in Fig. \ref{fig:Distribution_fnft}:
\begin{equation}
\label{ell_f_theta_xi}
\left\{
\begin{array}{lcl}
\langle f_{n} \rangle (\theta,\xi) &=& \langle f \rangle  \{ 1 + a_{fn}(\xi) \cos 2(\theta - \theta_{\sigma}) \}  \\
\langle f_{t} \rangle (\theta,\xi) &=& \langle f \rangle  a_{ft}(\xi) \sin 2(\theta - \theta_{\sigma}) , 
\end{array}
\right.
\end{equation}
where $a_{fn}(\xi)$ and $a_{ft}(\xi) $ are the amplitudes of normal and tangential force 
anisotropies in the $\xi$-networks. 
Notice that we have $\langle f_{t} \rangle=0 $ as a consequence of weak 
correlation between the branch vectors and contact forces as shown in 
Fig. \ref{fig:correlation_flr} and the balance of force moments. Morevover, 
the orthogonality between the normal and tangential forces implies that  
the peak value of $\langle f_{t} \rangle(\theta,\xi)$ occurs at an angle rotated by 
$\pi/4$ with respect to that of are rotated to  those of $\langle f_{n} \rangle (\theta,\xi) $.

As for the branch length vectors, the calculation of the anisotropy parameters $a_{fn}(\xi)$ and $a_{ft}(\xi) $ 
can be done by means of the following {\it force tensors} \cite{Radjai1998,Azema2010}: 
\begin{equation}
\label{Chi_tensors_xi}
\left\{
\begin{array}{lcl}
\chi^{fn}_{\alpha \beta}(\xi) &=& 
\frac{1}{\langle f \rangle} \int\limits_{0}^\pi  \langle f_{n} \rangle(\theta, \xi)  n_\alpha(\xi)  n_\beta(\xi) P_ \theta(\theta, \xi) d \theta,  \\ 
\chi^{ft}_{\alpha \beta}(\xi) &=& 
\frac{1}{\langle f \rangle} \int\limits_{0}^\pi  \langle f_{t} \rangle(\theta, \xi)  n_\alpha(\xi)  t_\beta(\xi) P_ \theta(\theta, \xi) d \theta.   
\end{array}
\right.
\end{equation}      
With these definitions, the following relationships can easily be established: 
\begin{eqnarray}
a_{fn}(\xi)&=& 2\frac{\chi^{fn}_{1}(\xi) - \chi^{fn}_{2}(\xi)}{ {\mbox tr} [\bm \chi^{fn}(\infty)]}       - a_c(\xi),  \\
a_{ft}(\xi)&=& 2\frac{\chi^{f}_{1}(\xi) - \chi^{f}_{2}(\xi)}{{\mbox tr} [\bm \chi^{f}(\infty)]}         - a_c(\xi)- a_{fn}(\xi),
\label{Aniso_values_xi}
\end{eqnarray}       
where  $\bm \chi^{f} =  \bm \chi^{fn} + \bm \chi^{ft}$ and
the indices $1$ and $2$ refer to the principal values of each tensor. 
By construction, we have ${\mbox tr} (\bm \chi^{f}) = \chi^{f}_{1} + \chi^{f}_{2} = \langle f \rangle$. 
The two partial force anisotropies $a_{fn}$ and $a_{ft}$ may take positive or 
negative values depending on the orientations $\theta_{fn}$ and $\theta_{ft}$ 
of the two tensors with respect to $\theta_\sigma$.   

\begin{figure}
\includegraphics[width=8cm]{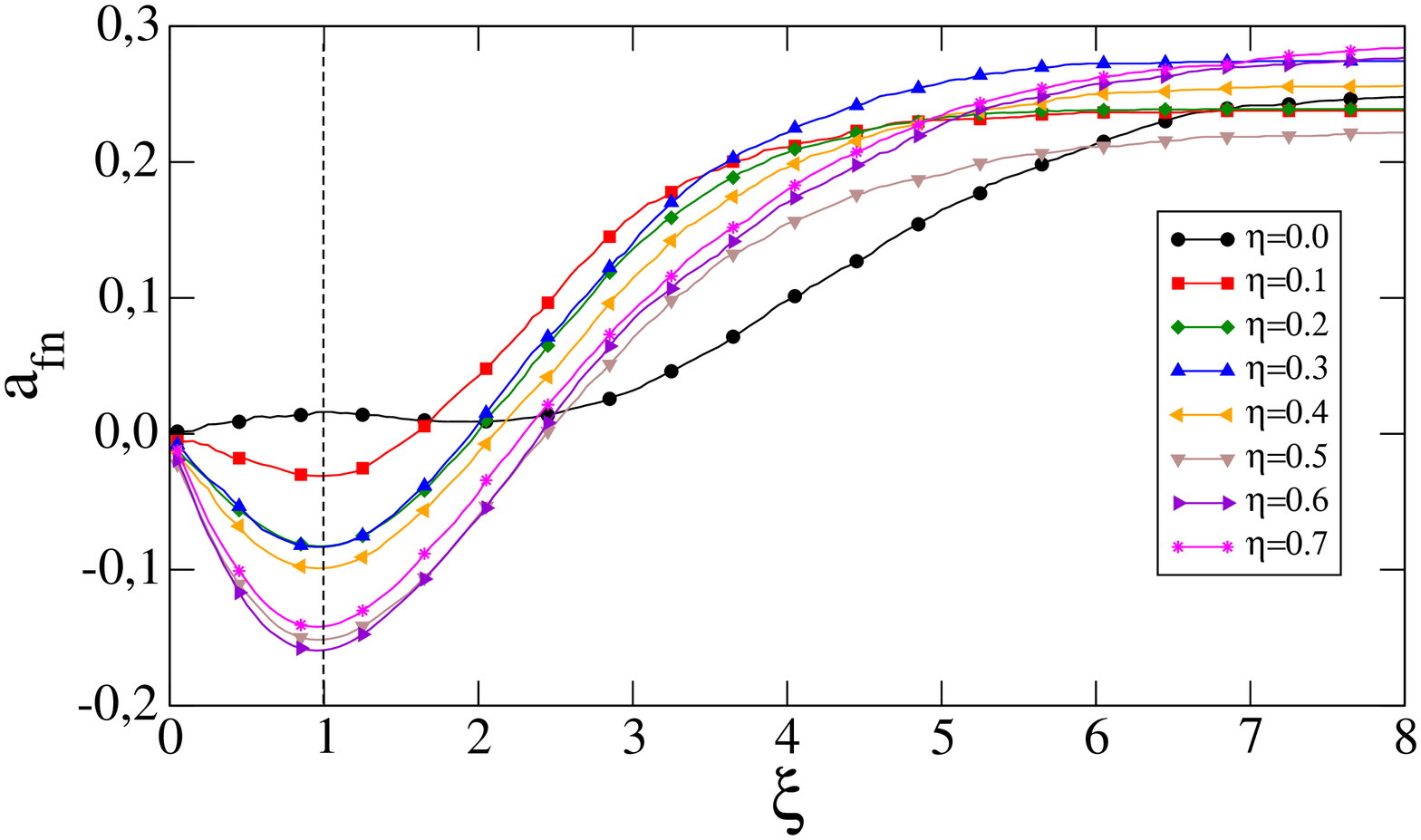}
\includegraphics[width=8cm]{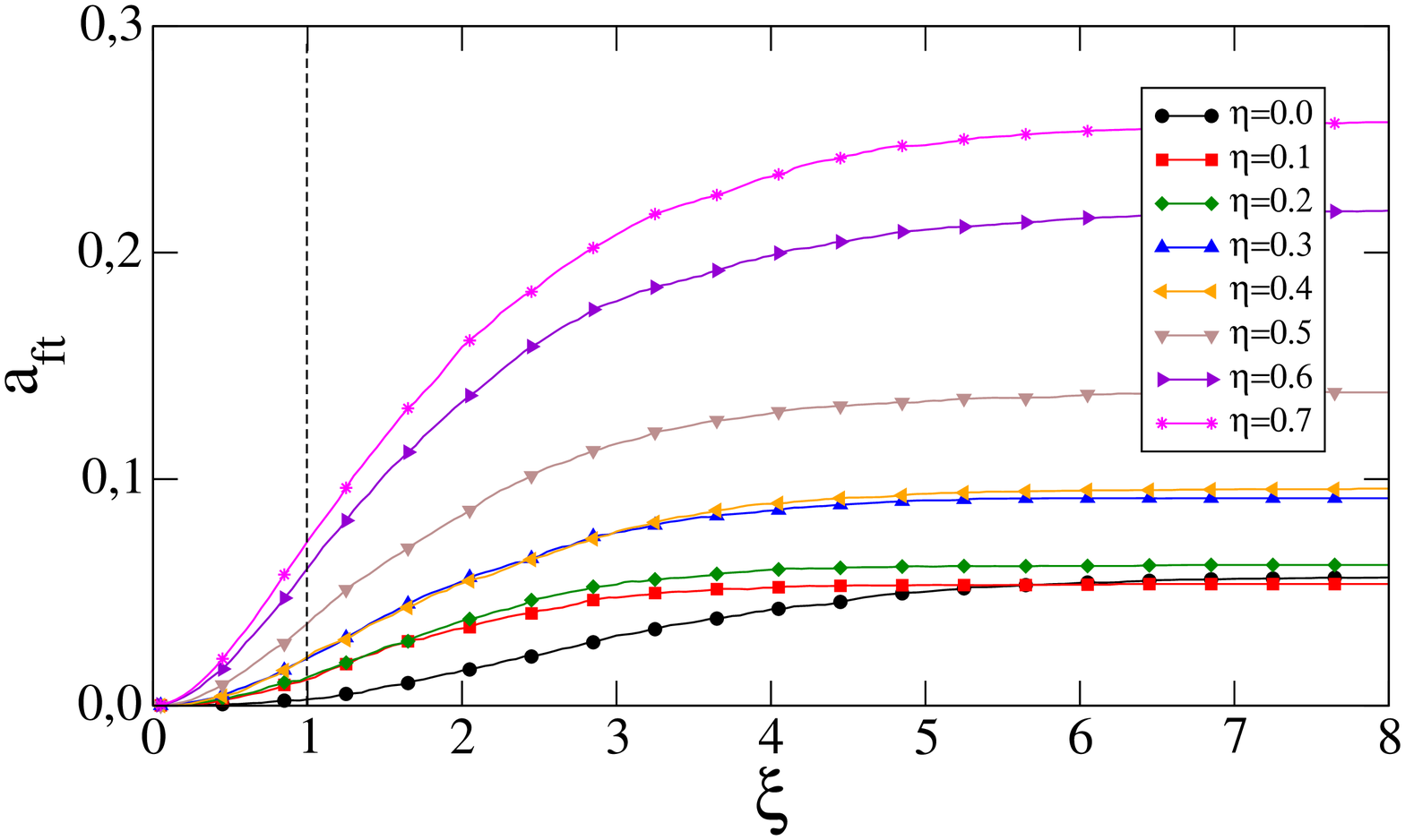}
\caption{Partial normal  and tangential  force anisotropies $a_{fn}$ and $a_{ft}$ 
as a function of force cutoff $\xi$ 
normalized by the mean force $\langle f \rangle$ for different values of  $\eta$.
\label{sec:fabric_force_transmission:afnt}}
\end{figure}

The normal and tangential force anisotropies are plotted in 
Fig.\ref{sec:fabric_force_transmission:afnt} as a function
of $\xi$ for all values of $\eta$. A remarkable feature 
of $a_{fn}(\xi)$ is that its value is negative in the weak network ($\xi<1$) 
for all elongated particles, i.e. for all values of $\eta$ in exception to $\eta=0$ 
where it remains positive for all $\xi$. Hence, the weak forces 
in a packing of elongated particles occur at contacts preferentially oriented 
orthogonally to the principal stress direction $\theta_\sigma$ whereas 
in a disk packing they are parallel. As we saw before, an inverse 
behavior occurs for the contact anisotropies, i.e. the weak contacts 
in the packings of elongated particles are parallel to the principal 
stress direction and orthogonal for the disk packings. $a_{fn}(\xi)$ increases in 
absolute value as $\xi$ increases and passes by a peak at 
exactly $\xi=1$, then declines as more contacts from the strong 
network with a positive contribution to the anisotropy 
are included in the $\xi$-network. At larger values (beyond $\xi \simeq 2$ for nearly all 
values of $\eta$), $a_{fn}(\xi)$ becomes positive as the strong forces tend to be 
parallel to the principal stress direction. This unmonotonic 
behavior of the partial force anisotropies for 
the elongated particles and the partial contact anisotropies for the disk packings 
underlies  the differentiation between the weak and strong networks 
according to the values of the normal contact forces with respect to the 
mean force ($\xi=1$). The difference between the elongated particle packings 
and disk packings reflects the formation of side-side contacts oriented along 
the principal stress direction tending to capture 
the strong force chains.      
                         
The tangential force anisotropy $a_{ft}(\xi)$ is an increasing function of 
both $\xi$ and $\eta$. Its value is generally below  
$a_{fn}(\xi)$, but becomes comparable for the most elongated particles 
for which the friction mobilization plays a key role as discussed previously. 
This is plausible as the tangential force anisotropy represents 
friction mobilization at contacts oriented at $\pi/4$ with respect to the 
major principal stress direction.      

\subsection{Stress tensor}

The physical importance of geometrical and mechanical anisotropies becomes clear when 
it is considered in connection with the 
stress tensor. As shown by Eq. \ref{eq:sigma}, the stress tensor 
is a function of discrete microscopic parameters
attached to the contact network. It is also possible to attribute a  
stress tensor to each $\xi$-network by restricting the summation to 
the corresponding contacts:
\begin{equation}
{\bm \sigma(\xi) } =   \frac{1}{V}  \sum_{c \in V} f_{\alpha}^c(\xi) \ell_{\beta}^c(\xi).
\label{eq:sigma_xi}
\end{equation}
For sufficiently large systems,  
the dependence of volume averages on individual discrete
parameters vanishes \cite{Rothenburg1989,Azema2010} and 
the discrete sums can be replaced by integrals as follows:
\begin{equation}
\sigma_{\alpha\beta}(\xi) = n_c \int_\Omega   f_\alpha(\xi) \ell_\beta(\xi) \ P_{\ell f}(\xi) d{\bm f} \ d{\bm \ell},
\label{eqn:isig_xi}
\end{equation}
where $P_{\ell f}$ is the joint probability density of forces and branch 
vectors in the $\xi$-networks, 
$n_c$ is the number density of contacts for the whole system 
and $\Omega$ is the integration domain 
in the space $(\bm \ell, \bm f)$.

The integral appearing in Eq. (\ref{eqn:isig_xi}) can be reduced by 
integrating first with respect to  the forces and branch vector lengths.  
Considering the components of the forces and branch vectors in contact  
frames $(\bm n, \bm t)$, and neglecting 
the branch/force correlations 
(see Fig.\ref{fig:Flr_Correlation}), 
we get \cite{Rothenburg1989,Azema2009,Azema2010}:
\begin{eqnarray}
\sigma_{\alpha\beta}(\xi) &=&  n_c \int\limits_0^{\pi}  
\{ \langle \ell_{n} \rangle (\theta, \xi) \ n_\alpha (\theta, \xi) + \langle \ell_{t} \rangle (\theta, \xi) \ t_\beta (\theta, \xi) \} \nonumber \\
 & \times & 
\{ \langle f_{n} \rangle (\theta, \xi) \ n_\alpha (\theta, \xi) + \langle f_{t} \rangle (\theta, \xi) \ 
t_\beta (\theta, \xi) \}  \nonumber \\ 
& \times & P(\theta, \xi) \ d\theta. 
\label{eqn:isig2_xi}
\end{eqnarray}
The expression of the stress tensor by this equation  makes appear explicitly 
the average directional functions representing the fabric and force states. 

\begin{figure}
\includegraphics[width=8cm]{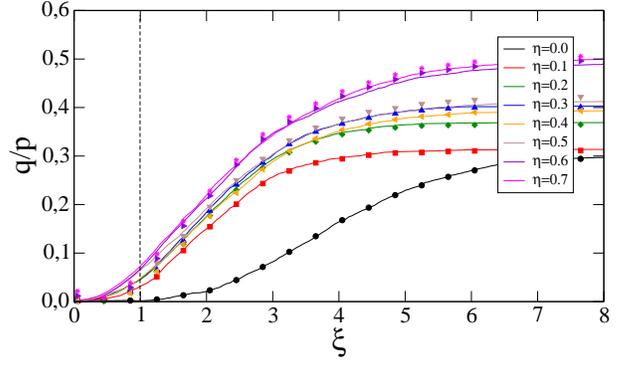}
\caption{Partial shear stress $q/p$ as a function of force cutoff $\xi$ 
for different values of  $\eta$ (plain line) together with approximation given
by Eq. \ref{q_p_aniso2_xi} (points).
\label{sec:fabric_force_transmission:qp}}
\end{figure}

Using the harmonic approximation introduced before, Eq. (\ref{eqn:isig2_xi}) can be 
integrated with respect to space direction $\theta$ and 
we get the following simple relation:      
\begin{equation}
\frac{q(\xi)}{p} \simeq 
\frac{1}{2} \{a_c(\xi) + a_{ln}(\xi) + a_{lt}(\xi) + a_{fn}(\xi) + a_{ft}(\xi) \},  
\label{q_p_aniso2_xi}
\end{equation}      
where the cross products among the anisotropy parameters have been neglected. 
This relation  expresses the normalized shear stress as a half-sum of texture and 
force anisotropies.    
Fig.\ref{sec:fabric_force_transmission:qp} displays the partial shear stress $q(\xi)/p$ as a function
of $\xi$ together with the approximation given by Eq. \ref{q_p_aniso2_xi}.
As we see, equation  (\ref{q_p_aniso2_xi}) provides an excellent fit to the data for all values 
of $\xi$ and $\eta$. 
Interestingly, $q(\xi<1)/p$ is zero for disk packings, implying that strong forces carry the
whole deviatoric load. The partial stress deviator $q(\xi=1)/p$ in the weak network increases 
slightly with $\eta$ but remains in all cases weak (below $0.1$). 
This transition reflects a qualitative change in the condition of 
local force balance in the presence of clusters as shown in Fig. \ref{fig:nematic}.
In other words, for these packings the weak 
network sustains also partially the deviatoric load applied to the system. 
The weak values of $q/p$ in the weak network is a consequence of 
the large positive value $a_c(\xi=1)=0.3$  which compensates 
the negative values of $a_{fn}(\xi=1)$, $a_{ln}(\xi=1)$ and $a_{lt}(\xi=1)$. 

\section{Summary}

In summary, using contacts dynamics simulations, we analyzed 
the granular texture and topology of forces chains in various packings composed of 
elongated particles under biaxial compression.
As compared to disk packings, the effect of particle elongation is to 
enhance the heterogeneity of the packings by the clustering of  
the particles according to their contact modes. In particular, the 
side/side contacts tend to capture strong force chains and be 
oriented orthogonally to the major principal stress direction. 
These features  
are reinforced as the particle elongation is increased.  
The probability densities of the normal forces become broader
with stronger force chains characterized by an exponential distribution
as in disks packings,
and with higher number of weak forces decreasing as a power law with the force.

An interesting finding of this work concerns the differentiation 
between the strong and weak force
networks for elongated particles. In contrast to disks packings, 
where the contacts in the weak network
are on the average perpendicular to the contacts in the strong network,
the contacts in a packing of elongated particles are, on the average, oriented
along the major principal stress direction both in the weak and strong networks.
But, the weak forces in the case of elongated particles show a negative anisotropy
in the sense that the average normal force in the weak network has its maximum
value in the contacts perpendicular to the strong network.
In other words, while in the disk packings the strong forces chains are 
propped by many weak lateral contact, for elongated particles
the strong force chains are laterally sustain by less contact but larger
weak forces.
A harmonic decomposition of the stress tensor shows, however, that for 
both disks and elongated particles, the compensating effects of force and contact
anisotropies lead to small shear stress deviator carried by the weak network.

Our simulation data indicate that the larger global shear strength of a packing of elongated particles
increases with elongation mainly due to the increase of friction mobilization and friction force 
anisotropy. The normal force anisotropy is large but nearly independent of elongation.
On the other hand, the correlation between contact forces and branch vectors joining particle centers 
reveal a sub-network of weak contacts with hight friction mobilization and small
branch vector length.

In conclusion, the packings of elongated particles in 2D reveal a nontrivial texture
allying the geometry of the particles with the preferred orientations of the contacts 
induced by shearing and equilibrium of particles. Some features are reminiscent of 
disk packings but are strongly modulated by the particle shape.
More work is underway to clarify the effect of particle shape by focusing on the 
local structures. On the other hand, many aspects of the packings 
analyzed in this paper are specific to two dimensions.
The side/side contacts in 3D between particles of spherocylindrical 
shape do not give rise to nematic ordering and the particle
rotations and forces moments play a major role
in the equilibrium of such particles. This point can only be analyzed 
by performing 3D simulations of large packings of sphero-cylinders
of varying elongation. However, since the class of side/side contacts 
controls to a large extent the specific behavior of elongated particles 
in 2D, we believe that similar features should occur in 3D for platy 
particles, which may give spontaneously rise to geometrical chains 
of face/face contacts.        
Such simulations require, however, much more computational 
effort. 

\bibliography{./azema}

\end{document}